# Effect of the spatial curvature on light bending and time delay in a curved Einstein-Straus–de Sitter spacetime

Mourad Guenouche[*]

*Laboratoire de Physique Théorique, Université Frères Mentouri-Constantine 1,
BP 325 route de Ain El Bey, 25017 Constantine, Algeria
and Université Abbès Laghrour de Khenchela, BP 1252, El Houria, Route de Constantine,
40004 Khenchela, Algeria*

A method of general applicability has been developed, whereby the null geodesic equations of the Einstein-Straus–de Sitter metric can be integrated simultaneously in terms of the curvature constant $k$. The purpose is to generalize the computation of light deflection and time delay by a spherical mass distribution. Assuming a flat Universe with most recent measurements of the Hubble constant $H_0$ and the cosmological constant $\Lambda$, five time delays between the four bright images of the lensed quasar SDSS J1004 + 4112 have been forecasted and compared to others in the field, $\Delta t_{DC} = (3250 \pm 64)$ days (8.90 yr), $\Delta t_{DA} = 2049^{+59}_{-58}$ days (5.61 yr), $\Delta t_{AC} = (1269 \pm 77)$ days (3.47 yr), $\Delta t_{BC} = 1176^{+78}_{-77}$ days (3.22 yr), and $\Delta t_{AB} = (93 \pm 70)$ days. This set of time delays constrains the galaxy cluster mass to be $M = (2.447 \pm 0.73) \times 10^{13} M_\odot$. In addition, we have reviewed the question of the possible contribution of a positive $\Lambda$ to reduce the light bending, and concluded that the changes are seemingly too small to be appreciable on cosmological scales. The same conclusion has been reached regarding the time delay. Having addressed the question of the effect of the spatial curvature in both closed and open Universe, we have found that the strong lensing is slightly affected by the expected small curvature density $\Omega_{k0}$ of the current Universe within its error bar $|\Omega_{k0}| \lessapprox 0.001$, in such a way that it may safely be neglected. However, it is only if $\Omega_{k0}$ gets quite larger that the effect being noticeable. While it is only theoretically possible for $\Omega_{k0}$ to be higher, it is worthwhile to stress that this should impact the light bending and time delay, causing them to decrease or increase depending upon whether the spatial curvature is positive or negative. Furthermore, one can infer that the observed light deflection and time delay independently, which are found to be significantly deviated from those of the flat Universe, may serve as a useful means to provide constraints on $\Omega_{k0}$, thus making the approach employed in this work more promising than others.

## I. INTRODUCTION

Einstein's general theory of relativity produces two famous testable evidences: the deflection of light near a massive body and the induced time delay. On the occasion of a total solar eclipse in 1919, Eddington had precisely measured the deflection angle due to the Sun, which matched perfectly with what general relativity predicts. The time delay is nothing more than the result of the light ray being longer than it would be without the Sun because of the deflection. This phenomenon is named Shapiro effect, after the person who discovered it between Earth and Mercury in 1964. In this case, the Sun, or any other star, is said to act as a microlens.

As for the large structures of the Universe, it is a matter of strong and weak lensing effects. Here, it is the strong lensing that we are concerned about. The light coming from background sources is strongly bent by a huge body, modifying the way these sources are seen from Earth and possibly giving rise to multiple images. The lensed quasar SDSS J1004 + 4112 (Sloan Digital Sky Survey) is a case in point. The number and the shape of the images depend on the lens geometry as well as on the source position. Moreover, according to observational measurements, the time delay between some widely separated images may spread over several years.

One of the main aims of this work is to generalize the computation of light deflection and time delay by a spherically symmetric mass distribution in the framework of the Einstein-Straus–de Sitter model (ESdS) [1–4], considering the three possible spatial curvatures of the Friedmann-Lemaître-Robertson-Walker (FLRW) metric: the positively curved space (closed Universe), the negatively curved space (open Universe), and the flat space (Euclidian universe). The choice of this swiss-cheese model is commonly justified by its relevance to the study of the gravitational lensing on cosmological scales. In such a model, the static Schwarzschild–de Sitter (SdS) or Kottler metric inside a vacuole (Schücking sphere) is glued to the homogeneous and isotropic expanding FLRW metric outside. In other words, to be consistent with the observational Hubble law, the model states that the gravitational lensing should only occur inside the vacuole to help explain why the cosmic expansion at larger scales, such as galaxies and clusters, is unobservable at smaller scales, such us planetary and atomic systems. Consequently, the model makes it possible to get rid of many assumptions adopted for simplicity within the SdS model, apart from the spherical symmetry which is of course impossible to overcome [5]. The observer and the source are allowed to be in movement with respect to the galaxy cluster lens as well as be in comovement with respect to a homogeneous isotropic dust including the other galaxy cluster masses, which mostly consists of cold dark matter (CDM). It is worth noting that the spatial curvature outside the vacuole is inherent in the entire FLRW Universe, not to be confused with the curvature of the SdS space-time due to the gravitational field caused by the mass distribution inside the vacuole.

---

[*]Contact author: guenouche_mourad@umc.edu.dz, guenouche_mourad@univ-khenchela.dz

Probing the precise shape of the universe remains an active area of research in cosmology. Observational current data, such as Planck measurements of the cosmic microwave background, indicate that the Universe is dominated by dark energy and CDM, and has a spatial curvature density very close to zero, $\Omega_{k0} = 0.0007 \pm 0.0019$ [6]. That is why most papers often focused on the simplest case of the flat space when dealing with strong lensing, in which the computation of light deflection and time delay is rather simple using features common to the Euclidean space [7–14]. In the nonflat spaces, this ceases to apply, as the photon no longer travels along straight lines outside the vacuole. In Ref. [15], we have already extended the analysis to address the same issue in the case where the expanding FLRW universe is rather positively curved than flat. To accomplish this, we have developed a method relying entirely on Einstein's postulates of general relativity, which are notoriously accurate in either flat or nonflat space-time. In this paper, we will generalize this method to cover a version of space with a negative curvature constant ($k = -1$), which has not yet been addressed. In this regard, a crucial question naturally arises as to how the light bending and the time delay are affected by the spatial curvature density at the present time and, to be consistent with the observation, whether one could ignore or not this effect in the limit of a weak spatial curvature density.

Not one of cosmologists claims that the time delay does not depend on the cosmological constant $\Lambda$, but they do not concur about the real effect on the deflection of light, despite the role that it plays in explaining the dark energy thought to be responsible for the cosmological expansion [16–31]. This question still unsettled is therefore worth further reexamining referring to recent measurements of cosmological parameters.

The paper is organized as follows. Section II is meant to construct the generalized ESdS metric by connecting the Schwarzschild–de Sitter metric with the FLRW one on the vacuole, from which we derive a formula relating the Schwarzschild coordinate time and that of Friedmann, in terms of a constant $k$ uniting both the positive/negative and zero curvature cases. From Sec. III, we get into the details to calculate the deflection of light and the time delay where only the implementation of a method based on integrating differential equations will be relevant to deal with the general case. We devote Sec. IV to validate our theoretical results through numerical applications[1] to the lensed quasar SDSS J1004 + 4112, already evoked above, by addressing each case separately. In the flat case, we provide estimates about the galaxy cluster mass, the deflection angle, and the time delays between the quasar images, and also test them against other recent predictions related to the subject. We also discuss how the light deflection is related to the cosmological constant. In the positively and negatively curved cases, we present a quantitative analysis in order to describe and discuss the evolution of the galaxy cluster mass, the deflection of light and the time delay in terms of the curvature density. In the last section, we end up with conclusions summarizing our main results and open questions offering new perspectives.

## II. JUNCTION CONDITIONS FOR GENERALIZED EINSTEIN-STRAUS–DE SITTER METRIC

The first section of this work is about constructing the generalized ESdS metric by linking the SdS and the generalized FLRW metrics, following the same treatment outlined in Refs. [8,10,15]. Let us denote by $(T, r, \theta, \varphi)$ the Schwarzschild coordinates and by $(t, \chi, \theta, \varphi)$ the Friedmann coordinates.

The static Schwarzschild–de Sitter metric takes the form

$$ds^2_{SdS} = B(r)dT^2 - B(r)^{-1}dr^2 - r^2 d\omega^2,$$
$$B(r) := 1 - \frac{r_{Schw}}{r} - \frac{\Lambda}{3}r^2, \qquad d\omega^2 = d\theta^2 + \sin^2\theta d\varphi^2, \quad (1)$$

inside a vacuole of radius $r_{Schü}$, usually named Schücking radius, $(T), \ r \leq r_{Schü}$, surrounding a spherical mass distribution $M$, with $r_{\text{Schw}} := 2GM$ the Schwarzschild radius. The generalized dynamic FLRW metric,

$$ds^2_{\text{FLRW}} = dt^2 - a(t)^2[d\chi^2 + \mathcal{S}(\chi)^2 d\omega^2], \quad (2)$$

describes the space-time geometry outside the vacuole, $\chi \geq \chi_{\text{Schü}}$, where we define the sine-like function $\mathcal{S}(\chi)$ to facilitate handling all three spatial curvatures simultaneously,

$$\mathcal{S}(\chi) := k^{-1/2} \sin(k^{1/2}\chi)$$
$$= \begin{cases} \sin\chi & k = +1 \text{ (positively curved space)} \\ \chi & k = 0 \text{ (flat space)} \\ \sinh\chi & k = -1 \text{ (negatively curved space)} \end{cases}, \quad (3)$$

and its cosinelike and tangentlike analogs, $\mathcal{C}(\chi) := \sqrt{1 - k\mathcal{S}(\chi)^2}$ and $\mathcal{T}(\chi) := \mathcal{S}(\chi)/\mathcal{C}(\chi)$. The evolution of the scale factor $a(t)$ over time is governed by the first order Friedmann equation

$$\frac{da}{dt} = aH(a), \quad H(a) := \sqrt{\frac{A}{a^3} - \frac{k}{a^2} + \frac{\Lambda}{3}}, \quad (4)$$

where the function $H(a)$ is the hubble parameter and $A$ is a constant coming from the energy conservation law relative to a nonrelativistic matter-dominated Universe characterized by a pressureless dust with density $\rho$, $3A := 8\pi G\rho a^3$. Then, $A = (H_0^2 - \Lambda/3)a_0^3 + ka_0$ obtained by writing (4) in the present time, where $H_0$ is called the hubble constant and $a_0$ is the scale factor at the present time. It is customary to rearrange the Friedmann equation in the usual standard form in terms of three density parameters: the matter density $\Omega_\rho$, the spatial curvature density $\Omega_k$, and the dark energy (or vacuum) density $\Omega_\Lambda$ as

$$1 = \Omega_\rho + \Omega_k + \Omega_\Lambda, \quad \Omega_\rho = \frac{A}{H(a)^2 a^3},$$
$$\Omega_k = \frac{-k}{H(a)^2 a^2}, \quad \Omega_\Lambda = \frac{\Lambda}{3H(a)^2}, \quad (5)$$

with $A := H_0^2 a_0^3(1 - \Omega_{k0} - \Omega_{\Lambda 0})$ rewritten in terms of the present spatial curvature density $\Omega_{k0}$ and the present dark energy density $\Omega_{\Lambda 0}$.

---
[1]All numerical assessments are performed using Wolfram *Mathematica* 11.

The two solutions are connected on the vacuole under the matching condition

$$r_{\text{Schü}}(T) := a(t)\mathcal{S}_{\text{Schü}}, \quad (6)$$

with $\mathcal{S}_{\text{Schü}} := \mathcal{S}(\chi_{\text{Schü}})$. Then, using the fact that $M$, $\rho$, and $r_{\text{Schü}}$ are related by $3M = 4\pi r^3_{\text{Schü}}\rho$, the constant Schücking radius can be expressed in terms of $A$ and $r_{\text{Schw}}$ as

$$\mathcal{S}_{\text{Schü}} = \left(\frac{r_{\text{Schw}}}{A}\right)^{\frac{1}{3}}. \quad (7)$$

So we have on the vacuole

$$B_{\text{Schü}} := B(r_{\text{Schü}}) = 1 - \left(\frac{A}{a} + \frac{\Lambda}{3}a^2\right)\mathcal{S}^2_{\text{Schü}}. \quad (8)$$

We will transform the Schwarzschild coordinates $(T, r)$ and the Friedmann coordinates $(t, \chi)$ into the new coordinates $(b, r)$. In this new coordinates, the SdS metric can recast as

$$ds^2_{\text{SdS}} = B(r)\Psi(b)^2 db^2 - \frac{1}{B(r)}dr^2 - r^2 d\omega^2, \quad (9)$$

by introducing a function $\Psi(b)$ defined as $\Psi(b) := dT/db$. As for the FLRW metric, an intermediate transformation to the coordinates $(a, \chi)$ must first be applied using the Friedmann equation (4), before transforming to the new coordinates $(b, r)$ by using

$$a := \Phi(b, r), \quad \mathcal{S}(\chi) := \frac{r}{\Phi(b, r)}. \quad (10)$$

We obtain, after getting rid of mixed terms,

$$ds^2_{\text{FLRW}} = \left(\frac{\partial\Phi}{\partial b}\right)^2 \frac{1 - [H(\Phi)^2 + k\Phi^{-2}]r^2}{H(\Phi)^2(\Phi^2 - kr^2)} db^2$$
$$- \frac{1}{1 - [H(\Phi)^2 + k\Phi^{-2}]r^2} dr^2 - r^2 d\omega^2, \quad (11)$$

with the boundary condition

$$a = b = \Phi(b, b\mathcal{S}_{\text{Schü}}), \quad (12)$$

that at the Schücking radius, old and new time coordinates coincide. We get, by differentiating with respect to $b$,

$$\left.\frac{\partial\Phi}{\partial b}\right|_{\text{Schü}} = 1 - \left.\frac{\partial\Phi}{\partial r}\right|_{\text{Schü}} \mathcal{S}_{\text{Schü}} = \frac{\mathcal{C}^2_{\text{Schü}}}{B_{\text{Schü}}(b)}, \quad (13)$$

where $\mathcal{C}_{\text{Schü}} := \mathcal{C}(\chi_{\text{Schü}})$, and $\partial\Phi/\partial r$ is given by

$$\frac{\partial\Phi}{\partial r} = -\frac{\Phi H(\Phi)^2 r}{1 - [H(\Phi)^2 + k\Phi^{-2}]r^2}. \quad (14)$$

The continuity condition is insured by equating the SdS metric components with those of FLRW at the Schücking radius. This requires

$$\Psi(b) = \frac{\mathcal{C}_{\text{Schü}}}{bH(b)B_{\text{Schü}}(b)}. \quad (15)$$

Now, the use of the chain rule allows us to calculate the Jacobian of the transformation from the Schwarzschild coordinates to the Friedmann coordinates at the Schücking radius, i.e.,

$$\left.\frac{\partial t}{\partial T}\right|_{\text{Schü}} = \mathcal{C}_{\text{Schü}}, \left.\frac{\partial t}{\partial r}\right|_{\text{Schü}} = -\frac{aH(a)\mathcal{S}_{\text{Schü}}}{B_{\text{Schü}}},$$
$$\left.\frac{\partial \chi}{\partial T}\right|_{\text{Schü}} = -H(a)\mathcal{S}_{\text{Schü}}, \left.\frac{\partial \chi}{\partial r}\right|_{\text{Schü}} = \frac{\mathcal{C}_{\text{Schü}}}{aB_{\text{Schü}}}. \quad (16)$$

Finally, considering the parametrized curve, $T = p$, $r = b\mathcal{S}_{\text{Schü}}$ ($\theta = \pi/2$, $\varphi = 0$) and calculating its 4-velocity, one can easily relate the Schwarzschild coordinate time $T$ to that of Friedmann $t$ on the vacuole. The resulting relationship is

$$\left.\frac{\mathrm{d}t}{\mathrm{d}T}\right|_{\text{Schü}} = \frac{\mathrm{d}t}{\mathrm{d}p}\frac{\mathrm{d}p}{\mathrm{d}T} = \frac{B_{\text{Schü}}}{\mathcal{C}_{\text{Schü}}}, \quad (17)$$

which ensure smooth passage of photons through the boundary between the SdS and FLRW geometries and vice versa.

### III. DEFLECTION OF LIGHT AND TIME DELAY

We suggest to study a typical scenario, as shown in Fig. 1, in which two photons are emitted by a source S at times $t_{\text{S}}$ and $t'_{\text{S}}$, enter the vacuole on one side at $t_{\text{SchüS}}$ and $t'_{\text{SchüS}}$, exit it on the other side at $t_{\text{SchüE}}$ and $t'_{\text{SchüE}}$, and eventually reach the Earth E at the same time $t_{\text{E}} = t'_{\text{E}} = 0$, where the final FLRW-type conditions are well known. It seems therefore more practical to carry out the integration of null geodesic equations backward in time, from the Earth to the vacuole, inside the vacuole, and from the vacuole to the source. Meanwhile, the use of the Jacobian transformation as well as its inverse are indispensable to convert the initial FLRW-type conditions to the final SdS-type conditions on the vacuole in front of the Earth, and the initial SdS-type conditions to the final FLRW-type conditions on the vacuole in front of the source. An added advantage of such a situation is that the time delay between both photons—difference between their total travel times from the emission times on the source to the receipt times being synchronized on Earth—will be simply expressed as $\Delta t \coloneqq \Delta t_{\text{S}} \coloneqq t_{\text{S}} - t'_{\text{S}}$. We denote by $\alpha$ and $\alpha'$ the angles that the two photons make upon receipt on Earth with the

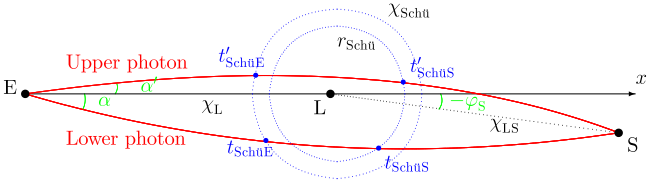

FIG. 1. Two photons emitted by a source S, bent inside the vacuole and finally received at Earth E. The trajectories outside the vacuole converge in the case of a spherical geometry of positive curvature, while they diverge in the case of a hyperbolic geometry of negative curvature. In a flat geometry, they are straight lines. We have limited ourselves to the spherical case to avoid overloading the figure.

Earth-lens axis, and by $r_{\text{P}}$ and $r'_{\text{P}}$, the minimum approach distances (perilens) at which they get deflected by the lens L.

It is important to note that even though the Friedmann equation (4) can be analytically solved in the case of nonzero curvature, with the aid of elliptic integrals of the second and third kinds, to provide the cosmic time $t(a)$ and the inverse function for the scale factor $a(t)$, which is still unknown in terms of the curvature constant $k$, unless in the case of a flat space $k = 0$ [10]. Instead, one can opt to solve the problem by means of the numerical integration method.

The numerical integration of the radial null geodesic equation

$$\mathrm{d}\chi = -\frac{\mathrm{d}t}{a(t)} \quad (18)$$

is also necessary to calculate the Earth-lens and the Earth-source geodesic distances, $\chi_{\text{L}}$ and $\chi_{\text{S}}$, where only the minus sign must be retained since the origin is defined on Earth, $\chi(t = 0) = 0$, leading to an increasing geodesic distance $\chi(t)$ over time. But, this equation involves $a(t)$, which is only analytically known in the case $k = 0$. In response to such a problem, one could introduce the Hubble parameter $H(a)$ and the cosmological redshift $z$ via the Friedmann equation (4) and the well-known formula $1 + z = a_0/a$ in order to integrate with respect to the scale factor,

$$\chi(z) = \int_{\frac{a_0}{1+z}}^{a_0} \frac{\mathrm{d}a}{a^2 H(a)}, \quad (19)$$

or with respect to $z$,

$$\chi(z) = \int_z^0 \frac{\mathrm{d}z}{a_0 H(z)},$$
$$H(z) = H_0\sqrt{\Omega_{\rho 0}(1+z)^3 + \Omega_{k0}(1+z)^2 + \Omega_{\Lambda 0}}. \quad (20)$$

Hence, $\chi_{\text{L}}$ and $\chi_{\text{S}}$ will be calculated for any given value of $z_{\text{L}}$ and $z_{\text{S}}$; $\chi_{\text{L}} \coloneqq \chi(z_{\text{L}})$ and $\chi_{\text{S}} \coloneqq \chi(z_{\text{S}})$. Concerning the geodesic distance that joins the lens to the source $\chi_{\text{LS}}$, it can be approximated as being aligned with $\chi_{\text{L}}$,

$$\chi_{\text{LS}} \simeq \chi_{\text{S}} - \chi_{\text{L}}, \quad (21)$$

due to the small $-\varphi_{\text{S}}$ of the order of a few arc seconds ($\sim 10^{-5}$).

#### A. Integration of null geodesic equations between the Earth and the vacuole

In this region, where the generalized FLRW metric (2) prevails, two geodesic equations are sufficient to describe the photon trajectory in the equatorial plane $\theta = \pi/2$,

$$\ddot{t}\ddot{t} + a\dot{a}[\dot{\chi}^2 + \mathcal{S}(\chi)^2\dot{\varphi}^2] = 0, \quad (22)$$

$$\frac{1}{2}\frac{\ddot{\varphi}}{\dot{\varphi}} + \frac{\dot{a}}{a} + \frac{\dot{\chi}}{\mathcal{T}(\chi)} = 0, \qquad (23)$$

where the dot denotes differentiation with respect to $p$, an affine parameter other than $s$ since the photon moves along null geodesics $ds = 0$.

The final FLRW-type conditions of the upper photon upon arrival on Earth ($t = 0, \chi = \chi_L, \varphi = \pi$) are given by its 4-velocity

$$\dot{t} = 1, \qquad \dot{\chi} = \frac{\cos\beta'}{a_0}, \qquad \dot{\varphi} = \frac{\sin\beta'}{a_0 \mathcal{S}_L}, \qquad (24)$$

where we have used the fact that the physical angle $\alpha'$ coincides with the coordinate angle $\arctan(\mathcal{T}_L |d\varphi/d\chi|_{\chi_L})$, and defined the constant $\beta'$ by $\tan\beta' := \mathcal{C}_L \tan\alpha'$, with $\mathcal{S}_L := \mathcal{S}(\chi_L)$, $\mathcal{C}_L := \mathcal{C}(\chi_L)$, and $\mathcal{T}_L := \mathcal{T}(\chi_L)$. These final FLRW-type conditions permit a straightforward integration of (22) and (23) to get

$$\dot{t} = \frac{a_0}{a(t)}, \qquad \dot{\varphi} = \frac{a_0 \mathcal{S}'_{PE}}{a^2 \mathcal{S}(\chi)^2}, \qquad \varphi = \pi - \arcsin\frac{\mathcal{T}'_{PE}}{\mathcal{T}(\chi)} + \beta', \qquad (25)$$

where the perilens $\chi'_{PE}$ is defined such that $\mathcal{S}'_{PE}, \mathcal{C}'_{PE}$, and $\mathcal{T}'_{PE}$ are related to each other by $\mathcal{S}'_{PE} := \mathcal{S}(\chi'_{PE}) := \mathcal{S}_L \sin\beta'$, $\mathcal{C}'_{PE} := \mathcal{C}(\chi'_{PE}) := \sqrt{\cos^2\beta' + \mathcal{C}_L^2 \sin^2\beta'}$, and $\mathcal{T}'_{PE} := \mathcal{T}(\chi'_{PE}) = \mathcal{S}'_{PE}/\mathcal{C}'_{PE}$.

Now, we are going to follow the approach originally proposed in Ref. [15], in order to determine the scale factor $a'_{SchüE}$ and its corresponding time $t'_{SchüE}$, at which the upper photon emerges from the vacuole. What we know up to now are the constant Schücking radius $\chi_{Schü}$ and its corresponding polar angle $\varphi'_{SchüE} := \varphi(\chi_{Schü})$ via the relationship between $\chi$ and $\varphi$ (25). So the idea is to relate one of them with time, $\chi(t)$ or $\varphi(t)$. Removing the affine parameter between $\dot{t}$ and $\dot{\varphi}$ in (25) and injecting the result into the generalized FLRW metric (2) for a photon, we get immediately

$$\frac{dt}{a(t)} = \frac{d\chi}{g(\chi)}, \qquad g(\chi) := \sqrt{1 - \frac{\mathcal{S}'^2_{PE}}{\mathcal{S}(\chi)^2}}, \qquad (26)$$

where we have taken into account that $\chi(t)$ is a decreasing function of time in this region. The left-hand side of this equation comprises $a(t)$ which we fail to know in the general case, as said before. So we should instead integrate with respect to the scale factor by inserting the Hubble parameter $H(a)$ through the Friedmann equation (4). The integration of the right-hand side can be analytically carried out to provide the function $-k^{-1/2} \arcsin[\mathcal{C}(\chi)/\mathcal{C}'_{PE}]$ and evaluated between the vacuole and the Earth to provide the geodesic distance traveled by the photon. The result is

$$\int_{a'_{SchüE}}^{a_0} \frac{da}{a^2 H(a)} = k^{-1/2} \left( \arcsin\frac{\mathcal{C}_{Schü}}{\mathcal{C}'_{PE}} - \arcsin\frac{\mathcal{C}_L}{\mathcal{C}'_{PE}} \right), \qquad (27)$$

which enables us to obtain $a'_{SchüE}$ by numerical integration. The corresponding value of $t'_{SchüE}$ is then calculated by numerical integration of the Friedmann equation, i.e.,

$$t'_{SchüE} = \int_{a_0}^{a'_{SchüE}} \frac{da}{aH(a)}, \qquad (28)$$

if one is interested in that. Let us note that the right-hand side of (27) reduces to

$$\text{rhs}(k = 0) = \sqrt{\chi_L^2 - \chi'^2_{PE}} - \sqrt{\chi^2_{Schü} - \chi'^2_{PE}}, \qquad (29)$$

equivalent to the result of Refs. [8,10] in the flat case $k = 0$, and to the result of Ref. [15] in the closed case $k = +1$. In the hyperbolic case $k = -1$, it reduces to

$$\text{rhs}(k = -1) = \text{arccosh}\frac{\cosh\chi_L}{\cosh\chi'_{PE}} - \text{arccosh}\frac{\cosh\chi_{Schü}}{\cosh\chi'_{PE}}. \qquad (30)$$

The time $t_{SchüE}$, at which the lower photon emerges from the vacuole, can be calculated in the same manner as before, using similar formulas. We just have to replace $\alpha'$ by $-\alpha$ as well as the polar angle $\pi$ by $-\pi$ on Earth.

However, we can proceed, in a way analogous to the flat and closed cases [10,15], to develop an analytical approximate expression of the time delay between the two photons at the exit from the vacuole, i.e., $\Delta t_{SchüE} := t_{SchüE} - t'_{SchüE}$. Subtracting the right and the left-hand side of (27) from the ones which correspond to the lower photon, one gets

$$\int_{t'_{SchüE}}^{t_{SchüE}} \frac{dt}{a(t)} = k^{-1/2} \left( \arcsin\frac{\mathcal{C}_{Schü}}{\mathcal{C}_{PE}} - \arcsin\frac{\mathcal{C}_L}{\mathcal{C}_{PE}} - \frac{\mathcal{C}_{Schü}}{\mathcal{C}'_{PE}} + \arcsin\frac{\mathcal{C}_L}{\mathcal{C}'_{PE}} \right). \qquad (31)$$

The left-hand side of this equation can be approximated using the fact that the scale factor varies significantly only on cosmological timescales, i.e.,

$$\text{lhs} \simeq \frac{\Delta t_{SchüE}}{a'_{SchüE}}, \qquad (32)$$

while an expansion of the right-hand side to the second-to-leading order in $\mathcal{S}'_{PE}$ and $\mathcal{S}_{PE}$ ($\sim 10^{-6}$) or in $\alpha'$ and $\alpha$ ($\sim 10^{-5}$) can be carried out to give

$$\text{rhs} \simeq \frac{1}{2}(\mathcal{T}^{-1}_{Schü} - \mathcal{T}^{-1}_L)\mathcal{S}_L^2 \mathcal{C}_L^2 (\alpha'^2 - \alpha^2), \qquad (33)$$

where $\mathcal{T}_{\text{Schü}} := \mathcal{T}(\chi_{\text{Schü}})$. Equating the two sides, we get

$$\Delta t_{\text{SchüE}} \simeq \frac{1}{2} a'_{\text{SchüE}} (\mathcal{T}_{\text{Schü}}^{-1} - \mathcal{T}_{\text{L}}^{-1}) \mathcal{S}_{\text{L}}^2 \mathcal{C}_{\text{L}}^2 (\alpha'^2 - \alpha^2). \tag{34}$$

Particularly, this is negative for $\alpha' < \alpha$, meaning that the lower photon precedes the upper one at the exit from the vacuole before they are received all together on Earth.

One can easily recover the results of the flat and the closed ESdS models in Refs. [10,15], taking $k = 0$ and $k = +1$, as well as the result of the open ESdS model, taking $k = -1$,

$$\Delta t_{\text{SchüE}} \simeq \frac{1}{2} a'_{\text{SchüE}} (\coth\chi_{\text{Schü}} - \coth\chi_{\text{L}}) \sinh^2\chi_{\text{L}} \cosh^2\chi_{\text{L}} (\alpha'^2 - \alpha^2). \tag{35}$$

### B. Integration of null geodesic equations inside the vacuole

At this step when the two photons permeate through the SdS space-time, the formulas that we are going to employ are extremely similar to those used in the flat and closed ESdS cases [10,15]. The only difference arises from the use of the relationship between the Schwarzschild time $T$ and the Friedmann time $t$ (17), which brings out the curvature constant $k$. We will therefore emphasize only what we need to calculate the time delay between the two photons at the entry into the vacuole represented by $\Delta t_{\text{SchüS}} := t_{\text{SchüS}} - t'_{\text{SchüS}}$.

Let us first determine the time it takes the upper photon for crossing the vacuole. Thanks to the relationship (17), this travel time can be expressed in terms of $t'_{\text{SchüE}}$ and $t'_{\text{SchüS}}$ as

$$T'_{\text{SchüE}} - T'_{\text{SchüS}} = \mathcal{C}_{\text{Schü}} \int_{t'_{\text{SchüS}}}^{t'_{\text{SchüE}}} \frac{\mathrm{d}t}{B_{\text{Schü}}(t)}. \tag{36}$$

The same quantity can be expressed otherwise by making use of the two well-known partially integrated geodesic equations of the SdS space-time (1),

$$\dot{T} = \frac{1}{B(r)}, \qquad \dot{r} = \pm\sqrt{1 - J'^2 \frac{B(r)}{r^2}}, \tag{37}$$

with $J'$ a constant of motion defined by $J' = r'_{\text{P}}/\sqrt{B(r'_{\text{P}})}$ and interpreted as an angular momentum per unit mass, $r'_{\text{P}}$ the perilens given approximately by $r'_{\text{P}} \simeq r'_{\text{SchüE}} \sin\gamma'_{\text{SdS}} - r_{\text{Schw}}/2$ [8], where $\gamma'_{\text{SdS}}$ denotes the smaller coordinate angle between the unoriented direction of the upper photon and the direction toward the lens, i.e., $\gamma'_{\text{SdS}} := \arctan|r'_{\text{SchüE}} \dot{\varphi}'_{\text{SchüE}}/\dot{r}'_{\text{SchüE}}|$, with $r'_{\text{SchüE}} := a'_{\text{SchüE}} \mathcal{S}_{\text{Schü}}$ (6), $\dot{\varphi}'_{\text{SchüE}} := \dot{\varphi}(r'_{\text{SchüE}})$ (25), and $\dot{r}'_{\text{SchüE}}$ calculated in terms of $\dot{t}_{\text{SchüE}}$ and $\dot{\chi}_{\text{SchüE}}$ using the inverse Jacobian of (16). Eliminating the affine parameter between $\dot{T}$ and $\dot{r}$ in (37), we find

$$\mathrm{d}T = \pm\frac{\mathrm{d}r}{v'(r)}, \qquad v'(r) := B(r)\sqrt{1 - J'^2\frac{B(r)}{r^2}}, \tag{38}$$

which gives

$$T'_{\text{SchüE}} - T'_{\text{SchüS}} = \left(\int_{r'_{\text{P}}}^{r'_{\text{SchüE}}} + \int_{r'_{\text{P}}}^{r'_{\text{SchüS}}}\right) \frac{\mathrm{d}r}{v'(r)}, \tag{39}$$

using the fact that $r$ decreases with time from $r'_{\text{SchüS}}$ to $r'_{\text{P}}$ while it increases from $r'_{\text{P}}$ to $r'_{\text{SchüE}}$. It follows from (36) and (39) that

$$\left(\int_{r'_{\text{P}}}^{a'_{\text{SchüE}} \mathcal{S}_{\text{Schü}}} + \int_{r'_{\text{P}}}^{a'_{\text{SchüS}} \mathcal{S}_{\text{Schü}}}\right) \frac{\mathrm{d}r}{v'(r)} = \mathcal{C}_{\text{Schü}} \int_{a'_{\text{SchüS}}}^{a'_{\text{SchüE}}} \frac{\mathrm{d}a}{aH(a) B_{\text{Schü}}(a)}, \tag{40}$$

where we have, as before, inserted the hubble parameter through the Friedmann equation in order to make possible the integration with respect to the scale factor, with $r'_{\text{SchüS}} := a'_{\text{SchüS}} \mathcal{S}_{\text{Schü}}$. This equation can be numerically solved to yield the value of $a'_{\text{SchüE}}$. If one is interested in $t'_{\text{SchüS}}$, then one can numerically integrate the Friedmann equation, i.e.,

$$t'_{\text{SchüS}} = \int_{a_0}^{a'_{\text{SchüS}}} \frac{\mathrm{d}a}{aH(a)}. \tag{41}$$

Similar formulas are applied leading to the calculation of the scale factor $a_{\text{SchüS}}$ when the lower photon immerses into the vacuole.

Let us now develop an analytical approximate expression for $\Delta t_{\text{SchüS}}$ following the same method described in Refs. [10,15]. We make use of the relationship (17) to write

$$\Delta T_{\text{SchüE}} - \Delta T_{\text{SchüS}} = \mathcal{C}_{\text{Schü}} \left( \int_{t'_{\text{SchüE}}}^{t_{\text{SchüE}}} - \int_{t'_{\text{SchüS}}}^{t_{\text{SchüS}}} \right) \frac{\mathrm{d}t}{B_{\text{Schü}}(t)}$$
$$\simeq \mathcal{C}_{\text{Schü}} \left( \frac{\Delta t_{\text{SchüE}}}{B'_{\text{SchüE}}} - \frac{\Delta t_{\text{SchüS}}}{B'_{\text{SchüS}}} \right), \tag{42}$$

taking into account that $B_{\text{Schü}}$ is only significant on cosmological timescales, with $\Delta T_{\text{SchüE}} := T_{\text{SchüE}} - T'_{\text{SchüE}}$, $\Delta T_{\text{SchüS}} := T_{\text{SchüS}} - T'_{\text{SchüS}}$, $B'_{\text{SchüE}} := B_{\text{Schü}}(t'_{\text{SchüE}})$, and $B'_{\text{SchüE}} := B_{\text{Schü}}(t'_{\text{SchüE}})$. In fact, this express the difference in the travel times between both photons inside the vacuole. This latter could be written in a different way (differently) by making use of (38) for the upper photon and its analogous formula for the lower one, i.e.,

$$\Delta T_{\text{SchüE}} - \Delta T_{\text{SchüS}} = \left( \int_{r'_{\text{SchüE}}}^{r_{\text{SchüE}}} + \int_{r'_{\text{SchüS}}}^{r_{\text{SchüS}}} \right) \frac{\mathrm{d}r}{v'(r)} - \Delta T_{\text{SdS}}, \tag{43}$$

where we have split up the integrals in such a way as to produce the following expression:

$$\Delta T_{\text{SdS}} := \left( \int_{r'_{\text{P}}}^{r'_{\text{SchüE}}} + \int_{r'_{\text{P}}}^{r'_{\text{SchüS}}} \right) \frac{\mathrm{d}r}{v'(r)} - \left( \int_{r_{\text{P}}}^{r'_{\text{SchüE}}} + \int_{r_{\text{P}}}^{r'_{\text{SchüS}}} \right) \frac{\mathrm{d}r}{v(r)}, \tag{44}$$

which can be compared to an expression already involved in the calculation of the time delay in the framework of the Schwarzschild–de Sitter solution [32], i.e.,

$$\Delta T_{\text{SdS}} \simeq \frac{1}{2}(r_{\text{P}}^2 - r_{\text{P}}'^2) \left( \frac{1}{r'_{\text{SchüE}}} + \frac{1}{r'_{\text{SchüS}}} \right) + 2r_{\text{Schw}} \ln \frac{r_{\text{P}}}{r'_{\text{P}}} - \frac{3}{8}\left(1 - \frac{r_{\text{P}}'^2}{r_{\text{P}}^2}\right)$$
$$\times \frac{r_{\text{Schw}}^2}{r_{\text{P}}'^2} \sqrt{\frac{3}{\Lambda}} \left[ \operatorname{arctanh}\left(\sqrt{\frac{\Lambda}{3}} r'_{\text{SchüE}}\right) + \operatorname{arctanh}\left(\sqrt{\frac{\Lambda}{3}} r'_{\text{SchüS}}\right) \right]. \tag{45}$$

Furthermore, since the lengths and timescales that we are dealing with are smaller than cosmological ones, the first term on the right-hand side of (43) can be approximated by

$$\left( \int_{r'_{\text{SchüE}}}^{r_{\text{SchüE}}} + \int_{r'_{\text{SchüS}}}^{r_{\text{SchüS}}} \right) \frac{\mathrm{d}r}{v'(r)} \simeq \frac{\Delta r_{\text{SchüE}}}{v'_{\text{SchüE}}} + \frac{\Delta r_{\text{SchüS}}}{v'_{\text{SchüS}}}$$
$$\simeq \left( \frac{\Delta a_{\text{SchüE}}}{v'_{\text{SchüE}}} + \frac{\Delta a_{\text{SchüS}}}{v'_{\text{SchüS}}} \right) \mathcal{S}_{\text{Schü}}$$
$$\simeq \frac{H'_{\text{SchüE}} r'_{\text{SchüE}} \Delta t_{\text{SchüE}}}{v'_{\text{SchüE}}} + \frac{H'_{\text{SchüS}} r'_{\text{SchüS}} \Delta t_{\text{SchüS}}}{v'_{\text{SchüS}}}, \tag{46}$$

where we have used the matching condition (6) and the Friedmann equation with $\Delta r_{\text{SchüE}} := r_{\text{SchüE}} - r'_{\text{SchüE}}$, $\Delta r_{\text{SchüS}} := r_{\text{SchüS}} - r'_{\text{SchüS}}$, $\Delta a_{\text{SchüE}} := a_{\text{SchüE}} - a'_{\text{SchüE}}$, $\Delta a_{\text{SchüS}} := a_{\text{SchüS}} - a'_{\text{SchüS}}$, $v'_{\text{SchüE}} := v'(r'_{\text{SchüE}})$, $v'_{\text{SchüS}} := v'(r'_{\text{SchüS}})$, $H'_{\text{SchüE}} := H(a'_{\text{SchüE}})$, and $H'_{\text{SchüS}} := H(a'_{\text{SchüS}})$. Using this together with (45), (43), and (42), we therefore get

$$\Delta t_{\text{SchüS}} \simeq \frac{\Delta T_{\text{SdS}} + (B'^{-1}_{\text{SchüE}} \mathcal{C}_{\text{Schü}} - H'_{\text{SchüE}} v'^{-1}_{\text{SchüE}} r'_{\text{SchüE}}) \Delta t_{\text{SchüE}}}{B'^{-1}_{\text{SchüS}} \mathcal{C}_{\text{Schü}} + v'^{-1}_{\text{SchüS}} H'_{\text{SchüS}} r'_{\text{SchüS}}}. \tag{47}$$

Of course, this result generates what has been obtained in the flat and closed ESdS models [10,15], as well as that of the hyperbolic ESdS model,

$$\Delta t_{\text{SchüS}} \simeq \frac{\Delta T_{\text{SdS}} + (B'^{-1}_{\text{SchüE}} \cosh \chi_{\text{Schü}} - H'_{\text{SchüE}} v'^{-1}_{\text{SchüE}} a'_{\text{SchüE}} \sinh \chi_{\text{Schü}}) \Delta t_{\text{SchüE}}}{B'^{-1}_{\text{SchüS}} \cosh \chi_{\text{Schü}} + v'^{-1}_{\text{SchüS}} H'_{\text{SchüS}} a'_{\text{SchüS}} \sinh \chi_{\text{Schü}}}. \tag{48}$$

Remarkably, $\Delta t_{\text{SchüS}}$ is positive if $\alpha' < \alpha$, contrary to $\Delta t_{\text{SchüE}}$. This will be explained later, once the total time delay $\Delta t$ have been calculated.

We close this section by calculating the polar angles $\varphi'_{\text{SchüS}}$ and $\varphi_{\text{SchüS}}$ needed for the next section, at which the upper and lower photons penetrate into the vacuole. We shall exploit the third well-known equation, which for the upper photon reads

$$d\varphi = \pm \frac{dr}{u'(r)}, \qquad u'(r) := r\sqrt{\frac{r^2}{r'^2_{\text{P}}} - 1} \sqrt{1 - \frac{r_{\text{Schw}}}{r'_{\text{P}}} \left( \frac{r'_{\text{P}}}{r} + \frac{1}{1 + r'_{\text{P}}/r} \right)}. \tag{49}$$

This results from eliminating the affine parameter between $\dot{r}$ in (37) and $\dot{\varphi}$ of the third well known partially integrated geodesic equation,

$$\dot{\varphi} = J'/r^2, \tag{50}$$

where the cosmological constant $\Lambda$ is incidentally erased. The integration of the above equation leads to

$$\varphi'_{\text{SchüE}} - \varphi'_{\text{SchüS}} = \left( \int_{r'_{\text{P}}}^{r'_{\text{SchüE}}} + \int_{r'_{\text{P}}}^{r'_{\text{SchüS}}} \right) \frac{dr}{u'(r)}, \tag{51}$$

taking into account that $\varphi$ increases when the upper photon approaches the lens as well as when it moves away. This gives, to the first leading order in the ratio $r_{\text{Schw}}/r'_{\text{P}}$,

$$\varphi'_{\text{SchüS}} \simeq \varphi'_{\text{SchüE}} - \pi + \arcsin \frac{r'_{\text{P}}}{r'_{\text{SchüE}}} + \arcsin \frac{r'_{\text{P}}}{r'_{\text{SchüS}}}$$
$$- \frac{r_{\text{Schw}}}{2r'_{\text{P}}} \left[ \left( 1 + \frac{1}{1 + r'_{\text{P}}/r'_{\text{SchüE}}} \right) \sqrt{1 - \frac{r'^2_{\text{P}}}{r'^2_{\text{SchüE}}}} \right.$$
$$\left. + \left( 1 + \frac{1}{1 + r'_{\text{P}}/r'_{\text{SchüS}}} \right) \sqrt{1 - \frac{r'^2_{\text{P}}}{r'^2_{\text{SchüS}}}} \right]. \tag{52}$$

In the same manner, we obtain a similar expression for the lower photon, where $\varphi_{\text{SchüS}}$ and $\varphi_{\text{SchüE}}$ differ from those of the upper one by a minus sign.

### C. Integration of null geodesic equations between the vacuole and the source

In this part of space-time, the motion of photons are ruled by the same geodesic equations of the FLRW metric, (22) and (23). To integrate them, we shall adopt the same technique previously developed in Sec. III A, taking into account the final FLRW-type conditions of the upper photon at the entry into the vacuole ($t = t'_{\text{SchüS}}$, $\chi = \chi_{\text{Schü}}, \varphi = \varphi'_{\text{SchüS}}$), i.e.,

$$\dot{t} = \dot{t}'_{\text{SchüS}}, \qquad \dot{\chi} = \dot{\chi}'_{\text{SchüS}}, \qquad \dot{\varphi} = \dot{\varphi}'_{\text{SchüS}}. \tag{53}$$

These final FLRW-type conditions are calculated by converting the initial SdS-type conditions ($\dot{T}'_{\text{SchüS}}, \dot{r}'_{\text{SchüS}}$) (37) using the Jacobian (16). We obtain

$$\dot{t} = \frac{E'}{a(t)}, \qquad \dot{\varphi} = \frac{J'}{a^2 \mathcal{S}(\chi)^2},$$
$$\varphi = \varphi'_{\text{SchüS}} + \arcsin \frac{\mathcal{T}'_{\text{PS}}}{\mathcal{T}(\chi)} - \gamma'_{\text{FLRW}}, \tag{54}$$

with $E'$ being a second constant of motion given by $E' := a'_{\text{SchüS}} \dot{t}'_{\text{SchüS}}$, involved, with $J'$ (37), in the definition of the perilens $\chi'_{\text{PS}}$ such that $\mathcal{S}'_{\text{PS}}, \mathcal{C}'_{\text{PS}}$, and $\mathcal{T}'_{\text{PS}}$ are related to each other by $\mathcal{S}'_{\text{PS}} := \mathcal{S}(\chi'_{\text{PS}}) := J'/E'$, $\mathcal{C}'_{\text{PS}} := \mathcal{C}(\chi'_{\text{PS}}) := \sqrt{1 - kJ'^2/E'^2}$, $\mathcal{T}'_{\text{PS}} := \mathcal{T}(\chi'_{\text{PS}}) = \mathcal{S}'_{\text{PS}}/\mathcal{C}'_{\text{PS}}$, and $\gamma'_{\text{FLRW}}$ is an angle defined by $\sin \gamma'_{\text{FLRW}} := \mathcal{T}'_{\text{PS}}/\mathcal{T}_{\text{Schü}}$. It is noteworthy that one could easily check that $\gamma'_{\text{FLRW}}$ represents the smaller physical angle between the unoriented direction of the upper photon and the direction toward the lens, i.e., $\gamma'_{\text{FLRW}} := \arctan(\mathcal{T}_{\text{Schü}} |d\varphi/d\chi|_{\text{SchüS}})$.

Hence, the inclination angle of the source $\varphi'_S$ corresponds to the geodesic distance between the lens to the source $\chi_{\text{LS}}$ (21), i.e.,

$$-\varphi'_S = -\varphi'_{\text{SchüS}} - \arcsin \frac{\mathcal{T}'_{\text{PS}}}{\mathcal{T}_{\text{LS}}} + \gamma'_{\text{FLRW}}, \tag{55}$$

with $\mathcal{T}_{\text{LS}} := \mathcal{T}(\chi_{\text{LS}}) = \mathcal{S}_{\text{LS}}/\mathcal{C}_{\text{LS}}$, $\mathcal{S}_{\text{LS}} := \mathcal{S}(\chi_{\text{LS}})$, and $\mathcal{C}_{\text{LS}} := \mathcal{C}(\chi_{\text{LS}})$. Following the same reasoning as for the upper photon, one arrives at similar formulas for the lower photon, where only the constant $J'$ differs from the first one by a minus sign. We restrict ourselves to giving the expression of the inclination angle $\varphi_S$,

$$-\varphi_S = -\varphi_{\text{SchüS}} + \arcsin\frac{\mathcal{T}_{\text{PS}}}{\mathcal{T}_{\text{LS}}} - \gamma_{\text{FLRW}}, \quad (56)$$

which must be equal to $\varphi'_S$, since both photons are emitted by the same source, where $\mathcal{T}_{\text{PS}} := \mathcal{T}(\chi_{\text{PS}}) = \mathcal{S}_{\text{PS}}/\mathcal{C}_{\text{PS}}$, $\mathcal{S}_{\text{PS}} := \mathcal{S}(\chi_{\text{PS}}) := J/E$, $\mathcal{C}_{\text{PS}} := \mathcal{C}(\chi_{\text{PS}}) := \sqrt{1 - kJ^2/E^2}$, $E := a_{\text{SchüS}} \dot{t}_{\text{SchüS}}$, and $\sin\gamma_{\text{FLRW}} := \mathcal{T}_{\text{PS}}/\mathcal{T}_{\text{Schü}}$.

However, these angles $\varphi'_S$ and $\varphi_S$ differ for randomly selected parameters. As explained in Ref. [15], of all the parameters involved, fitting the value of the lens mass turns out to be the only way to fulfil the required equality between them. Once the desired mass value is well established, we move to calculate the time delay thereafter.

Let us determine the scale factor $a'_S$ at the emission time $t'_S$ of the upper photon. Applying the same method followed in the Sec. III A, one finds easily an equation similar to (26) relating $\chi$ to the time, by using $\mathcal{S}'_{\text{PS}}$ instead of $\mathcal{S}'_{\text{PE}}$,

$$\frac{dt}{a(t)} = -\frac{d\chi}{h'(\chi)}, \quad h'(\chi) = \sqrt{1 - \frac{\mathcal{S}'^2_{\text{PS}}}{\mathcal{S}(\chi)^2}}, \quad (57)$$

where the minus sign ensures that $\chi$ decreases over time between the source and the vacuole. Similarly, integrating the right-hand side of the previous equation between $\chi_{\text{LS}}$ and $\chi_{\text{Schü}}$, one arrives at

$$\int_{a'_S}^{a'_{\text{SchüS}}} \frac{da}{a^2 H(a)} = k^{-1/2}\left(\arcsin\frac{\mathcal{C}_{\text{Schü}}}{\mathcal{C}'_{\text{PS}}} - \arcsin\frac{\mathcal{C}_{\text{LS}}}{\mathcal{C}'_{\text{PS}}}\right), \quad (58)$$

which, through the use of the Friedmann equation, can be numerically solved for obtaining $a'_S$. Then, if one is interested in $t'_S$, it suffices us to use the Friedmann equation, i.e.,

$$t'_S = \int_{a_0}^{a'_S} \frac{da}{aH(a)}. \quad (59)$$

Likewise, the scale factor $a_S$ and its corresponding emission time $t_S$ of the lower photon are calculated by similar formulas to (58) and (59), replacing $a'_{\text{SchüS}}$ by $a_{\text{SchüS}}$, and $\chi'_{\text{PS}}$ by $\chi_{\text{PS}}$.

Again, it is possible to proceed differently, such as in the flat and closed cases [10,15], by directly calculating the time delay through an approximate analytical expression, instead of calculating $a_S$ and $t_S$ separately. The idea consists of subtracting the two sides of (58) from their analogs of the lower photon and evaluating the integrals with respect to time. One gets

$$\int_{t'_{\text{SchüS}}}^{t_{\text{SchüS}}} \frac{dt}{a(t)} - \int_{t'_S}^{t_S} \frac{dt}{a(t)} = \arcsin\frac{\mathcal{C}_{\text{Schü}}}{\mathcal{C}'_{\text{PS}}} - \arcsin\frac{\mathcal{C}_{\text{LS}}}{\cos\chi'_{\text{PS}}}$$
$$- \arcsin\frac{\mathcal{C}_{\text{Schü}}}{\mathcal{C}_{\text{PS}}} + \arcsin\frac{\mathcal{C}\chi_{\text{LS}}}{\mathcal{C}_{\text{PS}}}. \quad (60)$$

As before, because the scale factor $a(t)$ varies noticeably only over cosmological timescales, the left-hand side of the above equation can be approximated as

$$\text{lhs} \simeq \frac{\Delta t_{\text{SchüS}}}{a'_{\text{SchüS}}} - \frac{\Delta t_S}{a'_S}. \quad (61)$$

TABLE I. Galaxy cluster mass $M$, position angle $-\varphi_S$, and time delay for the observed image pairs, $(A,B)$, $(A,C)$, $(D,A)$, $(B,C)$, and $(D,C)$ of the lensed quasar SDSS J1004 + 4112 in flat Einstein-Straus–de Sitter space-time ($k=0$). Bold values correspond to upper and lower limits.

| Images | $(\Omega_{\Lambda 0}, \alpha, \alpha')$ | $M[10^{13} M_\odot]$ | $-\varphi_S[\prime\prime]$ | $\Delta t[\text{days}]$ |
|---|---|---|---|---|
| $(A,B)$ | $(+,+,+)$ | **2.735** | 0.2196 | 94 |
| | $(+,-,+)$ | 2.709 | **0.3837** | **163** |
| | $(\pm 0, \pm 0, \pm 0)$ | 2.706 | 0.2211 | 93 |
| | $(-,+,-)$ | 2.702 | **0.0561** | **23** |
| | $(-,-,-)$ | **2.677** | 0.2225 | 93 |
| $(A,C)$ | $(+,+,+)$ | **3.137** | 2.7878 | 1277 |
| | $(+,-,+)$ | 3.108 | **2.9519** | **1346** |
| | $(\pm 0, \pm 0, \pm 0)$ | 3.106 | 2.8069 | 1269 |
| | $(-,+,-)$ | 3.103 | **2.6588** | **1192** |
| | $(-,-,-)$ | **3.074** | 2.8252 | 1260 |
| $(D,A)$ | $(+,+,+)$ | **1.764** | 5.9872 | 2066 |
| | $(+,-,+)$ | 1.738 | **6.1514** | **2108** |
| | $(\pm 0, \pm 0, \pm 0)$ | 1.740 | 6.0282 | 2049 |
| | $(-,+,-)$ | 1.742 | **5.9012** | **1991** |
| | $(-,-,-)$ | **1.717** | 6.0675 | 2032 |
| $(B,C)$ | $(+,+,+)$ | **3.177** | 2.5682 | 1184 |
| | $(+,-,+)$ | 3.147 | **2.7324** | **1254** |
| | $(\pm 0, \pm 0, \pm 0)$ | 3.145 | 2.5858 | 1176 |
| | $(-,+,-)$ | 3.143 | **2.4363** | **1099** |
| | $(-,-,-)$ | **3.113** | 2.6027 | 1168 |
| $(D,C)$ | $(+,+,+)$ | **2.049** | 8.7749 | 3276 |
| | $(+,-,+)$ | 2.019 | **8.9391** | **3314** |
| | $(\pm 0, \pm 0, \pm 0)$ | 2.023 | 8.8350 | 3250 |
| | $(-,+,-)$ | 2.026 | **8.7263** | **3186** |
| | $(-,-,-)$ | **1.997** | 8.8926 | 3224 |

TABLE II. Uncertainties in the galaxy cluster mass $M$, the position angle $-\varphi_S$ and the time delay of the observed images, $(A,B)$, $(B,C)$, $(A,C)$, $(D,A)$, and $(D,C)$ of the lensed quasar SDSS J1004 + 4112 in flat Einstein-Straus–de Sitter space-time ($k=0$), with $\Delta\alpha := \alpha - \alpha'$.

| Images | $\Delta\alpha(\prime\prime)$ | $M[10^{13} M_\odot]$ | $-\varphi_S[\prime\prime]$ | $\Delta t[\text{days}]$ |
|---|---|---|---|---|
| $(A,B)$ | 0.11 | $2.706 \pm 0.029$ | $0.2211^{+0.1626}_{-0.1650}$ | $93 \pm 70$ |
| $(B,C)$ | 1.25 | $3.145 \pm 0.032$ | $2.5858^{+0.1466}_{-0.1495}$ | $1176^{+78}_{-77}$ |
| $(A,C)$ | 1.36 | $3.106 \pm 0.032$ | $2.8069^{+0.1450}_{-0.1481}$ | $1269 \pm 77$ |
| $(D,A)$ | 2.92 | $1.740^{+0.024}_{-0.023}$ | $6.0282^{+0.1232}_{-0.1270}$ | $2049^{+59}_{-58}$ |
| $(D,C)$ | 4.28 | $2.023 \pm 0.026$ | $8.8350^{+0.1041}_{-0.1087}$ | $3250 \pm 64$ |

TABLE III. Variation of the galaxy cluster mass $M$, the position angle $-\varphi_S$, and the time delay $\Delta t$ versus the cosmological constant $\Lambda$ within its error bar $\pm 2.90744 \times 10^{-54}$ m$^{-2}$ for the observed image pairs $(A, B)$, $(B, C)$, $(A, C)$, $(D, A)$, and $(D, C)$ of the lensed quasar SDSS J1004 + 4112 in flat Einstein-Straus–de Sitter space-time ($k = 0$). The angles $\alpha$ and $\alpha'$ are fixed in their central values.

| Images | $(\Omega_{\Lambda 0}, \alpha, \alpha')$ | $\delta\Lambda[\%]$ | $M[10^{13} M_\odot]$ | $\delta M[\%]$ | $-\varphi_S['']$ | $\delta\varphi_S[\%]$ | $\Delta t$[days] | $\delta\Delta t[\%]$ |
|---|---|---|---|---|---|---|---|---|
| $(A, B)$ | $(-, \pm 0, \pm 0)(+, \pm 0, \pm 0)$ | 5.33897 | 2.702196 | 0.271264 | 0.22251 | −1.32403 | 93.0302 | 0.44038 |
| | | | 2.709526 | | 0.21957 | | 93.4399 | |
| $(B, C)$ | | | 3.140887 | 0.271134 | 2.60266 | −1.32411 | 1173.49 | 0.44219 |
| | | | 3.149403 | | 2.56820 | | 1178.68 | |
| $(A, C)$ | | | 3.101231 | 0.271150 | 2.82518 | −1.32398 | 1265.88 | 0.44253 |
| | | | 3.109640 | | 2.78777 | | 1271.49 | |
| $(D, A)$ | | | 1.737773 | 0.271580 | 6.06753 | −1.32362 | 2044.69 | 0.45774 |
| | | | 1.742492 | | 5.98721 | | 2054.05 | |
| $(D, C)$ | | | 2.019884 | 0.271501 | 8.89263 | −1.32376 | 3242.31 | 0.47344 |
| | | | 2.025368 | | 8.77491 | | 3257.66 | |

TABLE IV. Variation of the galaxy cluster mass $M$, the position angle $-\varphi_S$, and the time delay $\Delta t$ versus the present curvature density $\Omega_{k0}$ within the range $[-0.3, 0.3]$ for the observed image pair $(A, B)$ of the lensed quasar SDSS J1004 + 4112 in curved Einstein-Straus–de Sitter space-time ($k = \pm 1$). The angles $\alpha_A$ and $\alpha_B$ as well as the present dark energy density $\Omega_{\Lambda 0}$ are fixed in their central values.

| $(A, B)$ | $k = +1$ | | | $k = -1$ | | |
|---|---|---|---|---|---|---|
| $|\Omega_{k0}|$ | $M[10^{13} M_\odot]$ | $-\varphi_S['']$ | $\Delta t$[days] | $M[10^{13} M_\odot]$ | $-\varphi_S['']$ | $\Delta t$[days] |
| 0.0001 | 2.705905 | 0.221071 | 93.2337 | 2.706151 | 0.221158 | 93.2802 |
| 0.0002 | 2.705775 | 0.221069 | 93.2284 | 2.706268 | 0.221243 | 93.3220 |
| 0.0003 | 2.705646 | 0.221068 | 93.2236 | 2.706385 | 0.221328 | 93.3632 |
| 0.0004 | 2.705516 | 0.221066 | 93.2184 | 2.706501 | 0.221414 | 93.4049 |
| 0.0005 | 2.705386 | 0.221065 | 93.2131 | 2.706618 | 0.221499 | 93.4462 |
| 0.0006 | 2.705256 | 0.221064 | 93.2083 | 2.706735 | 0.221584 | 93.4879 |
| 0.0007 | 2.705128 | 0.221062 | 93.2030 | 2.706851 | 0.221670 | 93.5291 |
| 0.0008 | 2.704997 | 0.221061 | 93.1977 | 2.706968 | 0.221755 | 93.5709 |
| 0.0009 | 2.704867 | 0.221059 | 93.1925 | 2.707085 | 0.221841 | 93.6121 |
| 0.001 | 2.704738 | 0.221058 | 93.1877 | 2.707202 | 0.221926 | 93.6539 |
| 0.002 | 2.703442 | 0.221043 | 93.1359 | 2.708369 | 0.222782 | 94.0701 |
| 0.003 | 2.702147 | 0.221029 | 93.0847 | 2.709538 | 0.223639 | 94.4874 |
| 0.004 | 2.700854 | 0.221014 | 93.0335 | 2.710707 | 0.224499 | 94.9054 |
| 0.005 | 2.699562 | 0.221000 | 92.9829 | 2.711877 | 0.225360 | 95.3251 |
| 0.006 | 2.698271 | 0.220985 | 92.9317 | 2.713048 | 0.226223 | 95.7459 |
| 0.007 | 2.696981 | 0.220971 | 92.8811 | 2.714219 | 0.227088 | 96.1680 |
| 0.008 | 2.695692 | 0.220956 | 92.8299 | 2.715392 | 0.227955 | 96.5912 |
| 0.009 | 2.694405 | 0.220941 | 92.7786 | 2.716565 | 0.228824 | 97.0156 |
| 0.01 | 2.693118 | 0.220927 | 92.7280 | 2.717739 | 0.229694 | 97.4412 |
| 0.02 | 2.680320 | 0.220780 | 92.2230 | 2.729522 | 0.238505 | 101.763 |
| 0.03 | 2.667640 | 0.220632 | 91.7232 | 2.741385 | 0.247511 | 106.209 |
| 0.04 | 2.655076 | 0.220482 | 91.2288 | 2.753326 | 0.256720 | 110.785 |
| 0.05 | 2.642627 | 0.220332 | 90.7396 | 2.765345 | 0.266137 | 115.495 |
| 0.06 | 2.630291 | 0.22018 | 90.2563 | 2.777441 | 0.275773 | 120.344 |
| 0.07 | 2.618069 | 0.220028 | 89.7778 | 2.789613 | 0.285633 | 125.339 |
| 0.08 | 2.605957 | 0.219875 | 89.3045 | 2.801859 | 0.295727 | 130.486 |
| 0.09 | 2.593956 | 0.219720 | 88.8360 | 2.814178 | 0.306064 | 135.790 |
| 0.1 | 2.582064 | 0.219566 | 88.3727 | 2.826569 | 0.316654 | 141.258 |
| 0.2 | 2.468865 | 0.217984 | 83.9951 | 2.953836 | 0.438974 | 206.765 |
| 0.3 | 2.365248 | 0.216372 | 80.0388 | 3.083522 | 0.60307 | 300.287 |

TABLE V. Variation of the galaxy cluster mass $M$, the position angle $-\varphi_S$, and the time delay $\Delta t$ versus the present curvature density $\Omega_{k0}$ within the range $[-0.3, 0.3]$ for the observed image pair $(B, C)$ of the lensed quasar SDSS J1004 + 4112 in curved Einstein-Straus–de Sitter space-time ($k = \pm 1$). The angles $\alpha_B$ and $\alpha_C$ as well as the present dark energy density $\Omega_{\Lambda 0}$ are fixed in their central values.

| $(B, C)$ | $k = +1$ | | | $k = -1$ | | |
|---|---|---|---|---|---|---|
| $|\Omega_{k0}|$ | $M[10^{13} M_\odot]$ | $-\varphi_S['']$ | $\Delta t[\text{days}]$ | $M[10^{13} M_\odot]$ | $-\varphi_S['']$ | $\Delta t[\text{days}]$ |
| 0.0001 | 3.145197 | 2.58578 | 1176.06 | 3.145484 | 2.58589 | 1176.23 |
| 0.0002 | 3.145046 | 2.58576 | 1176.00 | 3.145620 | 2.58599 | 1176.33 |
| 0.0003 | 3.144895 | 2.58574 | 1175.93 | 3.145757 | 2.58608 | 1176.43 |
| 0.0004 | 3.144744 | 2.58572 | 1175.87 | 3.145894 | 2.58618 | 1176.53 |
| 0.0005 | 3.144594 | 2.58571 | 1175.81 | 3.146030 | 2.58628 | 1176.63 |
| 0.0006 | 3.144443 | 2.58569 | 1175.74 | 3.146167 | 2.58637 | 1176.73 |
| 0.0007 | 3.144292 | 2.58567 | 1175.68 | 3.146303 | 2.58647 | 1176.83 |
| 0.0008 | 3.144141 | 2.58566 | 1175.61 | 3.146440 | 2.58656 | 1176.93 |
| 0.0009 | 3.143990 | 2.58564 | 1175.55 | 3.146576 | 2.58666 | 1177.03 |
| 0.001 | 3.143840 | 2.58562 | 1175.48 | 3.146713 | 2.58676 | 1177.13 |
| 0.002 | 3.142334 | 2.58545 | 1174.83 | 3.148080 | 2.58773 | 1178.12 |
| 0.003 | 3.140829 | 2.58528 | 1174.19 | 3.149447 | 2.58869 | 1179.12 |
| 0.004 | 3.139326 | 2.58511 | 1173.54 | 3.150816 | 2.58967 | 1180.12 |
| 0.005 | 3.137824 | 2.58494 | 1172.89 | 3.152185 | 2.59064 | 1181.13 |
| 0.006 | 3.136324 | 2.58477 | 1172.25 | 3.153556 | 2.59161 | 1182.13 |
| 0.007 | 3.134825 | 2.58460 | 1171.61 | 3.154927 | 2.59259 | 1183.14 |
| 0.008 | 3.133327 | 2.58443 | 1170.96 | 3.156300 | 2.59356 | 1184.15 |
| 0.009 | 3.131830 | 2.58426 | 1170.32 | 3.157673 | 2.59454 | 1185.16 |
| 0.01 | 3.130335 | 2.58409 | 1169.68 | 3.159047 | 2.59552 | 1186.17 |
| 0.02 | 3.115461 | 2.58237 | 1163.30 | 3.172843 | 2.60542 | 1196.39 |
| 0.03 | 3.100723 | 2.58064 | 1156.98 | 3.186733 | 2.61548 | 1206.80 |
| 0.04 | 3.086121 | 2.57890 | 1150.74 | 3.200718 | 2.62571 | 1217.40 |
| 0.05 | 3.071652 | 2.57714 | 1144.56 | 3.214797 | 2.63613 | 1228.20 |
| 0.06 | 3.057315 | 2.57537 | 1138.45 | 3.228968 | 2.64672 | 1239.21 |
| 0.07 | 3.043109 | 2.57359 | 1132.41 | 3.243231 | 2.65751 | 1250.43 |
| 0.08 | 3.029033 | 2.57179 | 1126.43 | 3.257585 | 2.66850 | 1261.87 |
| 0.09 | 3.015085 | 2.56999 | 1120.52 | 3.272027 | 2.67969 | 1273.53 |
| 0.1 | 3.001262 | 2.56818 | 1114.66 | 3.286557 | 2.69109 | 1285.43 |
| 0.2 | 2.869695 | 2.54969 | 1059.38 | 3.436029 | 2.81872 | 1419.22 |
| 0.3 | 2.749263 | 2.53084 | 1009.41 | 3.589010 | 2.98000 | 1588.39 |

Concerning the right-hand side, one gets up to the second-to-leading order in $\mathcal{S}'_{PS}$ and $\mathcal{S}_{PS}$ ($\sim 10^{-6}$) or in $\varphi'_{SchüS} - \varphi_S$ and $|\varphi_{SchüS} - \varphi_S|$ ($\sim 10^{-4}$) on account of (55) and (56),

$$\text{lhs} \simeq \frac{1}{2} \frac{(\varphi_{SchüS} - \varphi_S)^2 - (\varphi'_{SchüS} - \varphi_S)^2}{\mathcal{T}^{-1}_{Schü} - \mathcal{T}^{-1}_{LS}}. \quad (62)$$

Equating the two sides, one finally arrives at the expression of the total time delay

$$\Delta t \simeq a'_S \left[ \frac{\Delta t_{SchüS}}{a'_{SchüS}} - \frac{1}{2} \frac{(\varphi_{SchüS} - \varphi_S)^2 - (\varphi'_{SchüS} - \varphi_S)^2}{\mathcal{T}^{-1}_{Schü} - \mathcal{T}^{-1}_{LS}} \right], \quad (63)$$

which is positive for $\alpha > \alpha'$. Obviously, this expression reproduces the time delays already calculated in Refs. [10,15] in the flat and closed ESdS models, as well as that of the hyperbolic ESdS model,

$$\Delta t \simeq a'_S \left[ \frac{\Delta t_{SchüS}}{a'_{SchüS}} - \frac{1}{2} \frac{(\varphi_{SchüS} - \varphi_S)^2 - (\varphi'_{SchüS} - \varphi_S)^2}{\coth(\chi_{Schü}) - \coth(\chi_{LS})} \right]. \quad (64)$$

Overall, we realize that the upper photon after being the first emitted by the source arrives at the vacuole earlier than the lower one, but in turn, comes out from it later in such a way as to be eventually received on Earth simultaneously with the lower photon. As a matter of fact, this happens because the upper photon, which takes the nearest path from the lens ($r'_P < r_P$), as illustrated in Fig. 1, more intensely experiences the space-time curvature around the lens mass.

For symmetry reasons, one can expect that the polar angle and the time delay cancel, $\varphi_S = 0$, $\Delta t = 0$, when $\alpha = \alpha'$, in which case the Earth and the lens, as well as the source are fully aligned to each other.

TABLE VI. Variation of the galaxy cluster mass $M$, the position angle $-\varphi_S$, and the time delay $\Delta t$ versus the present curvature density $\Omega_{k0}$ within the range $[-0.3, 0.3]$ for the observed image pair $(A, C)$ of the lensed quasar SDSS J1004 + 4112 in curved Einstein-Straus–de Sitter space-time ($k = \pm 1$). The angles $\alpha_A$ and $\alpha_C$ as well as the present dark energy density $\Omega_{\Lambda 0}$ are fixed in their central values.

| $(A, C)$ | $k = +1$ | | | $k = -1$ | | |
|---|---|---|---|---|---|---|
| $\|\Omega_{k0}\|$ | $M[10^{13} M_\odot]$ | $-\varphi_S['']$ | $\Delta t$[days] | $M[10^{13} M_\odot]$ | $-\varphi_S['']$ | $\Delta t$[days] |
| 0.0001 | 3.105487 | 2.80685 | 1268.67 | 3.105771 | 2.80696 | 1268.84 |
| 0.0002 | 3.105339 | 2.80683 | 1268.60 | 3.105906 | 2.80706 | 1268.94 |
| 0.0003 | 3.105190 | 2.80681 | 1268.53 | 3.106041 | 2.80715 | 1269.05 |
| 0.0004 | 3.105040 | 2.80679 | 1268.46 | 3.106176 | 2.80725 | 1269.15 |
| 0.0005 | 3.104892 | 2.80677 | 1268.39 | 3.106311 | 2.80735 | 1269.25 |
| 0.0006 | 3.104743 | 2.80675 | 1268.32 | 3.106446 | 2.80744 | 1269.36 |
| 0.0007 | 3.104594 | 2.80674 | 1268.25 | 3.106580 | 2.80754 | 1269.46 |
| 0.0008 | 3.104445 | 2.80672 | 1268.18 | 3.106716 | 2.80764 | 1269.56 |
| 0.0009 | 3.104296 | 2.80670 | 1268.11 | 3.106851 | 2.80773 | 1269.67 |
| 0.001 | 3.104147 | 2.80668 | 1268.04 | 3.106985 | 2.80783 | 1269.77 |
| 0.002 | 3.102661 | 2.8065 | 1267.34 | 3.108336 | 2.8088 | 1270.81 |
| 0.003 | 3.101175 | 2.80631 | 1266.64 | 3.109687 | 2.80976 | 1271.85 |
| 0.004 | 3.099691 | 2.80613 | 1265.94 | 3.111040 | 2.81073 | 1272.89 |
| 0.005 | 3.098208 | 2.80594 | 1265.25 | 3.112393 | 2.81170 | 1273.93 |
| 0.006 | 3.096726 | 2.80576 | 1264.55 | 3.113747 | 2.81268 | 1274.98 |
| 0.007 | 3.095246 | 2.80557 | 1263.85 | 3.115102 | 2.81365 | 1276.02 |
| 0.008 | 3.093767 | 2.80539 | 1263.16 | 3.116458 | 2.81463 | 1277.07 |
| 0.009 | 3.092290 | 2.80520 | 1262.47 | 3.117815 | 2.8156 | 1278.12 |
| 0.01 | 3.090814 | 2.80502 | 1261.77 | 3.119173 | 2.81658 | 1279.17 |
| 0.02 | 3.076127 | 2.80315 | 1254.89 | 3.132804 | 2.82646 | 1289.80 |
| 0.03 | 3.061575 | 2.80127 | 1248.08 | 3.146530 | 2.8365 | 1300.61 |
| 0.04 | 3.047157 | 2.79938 | 1241.34 | 3.160349 | 2.84671 | 1311.62 |
| 0.05 | 3.032870 | 2.79747 | 1234.68 | 3.174261 | 2.85709 | 1322.83 |
| 0.06 | 3.018715 | 2.79555 | 1228.08 | 3.188265 | 2.86765 | 1334.24 |
| 0.07 | 3.004688 | 2.79361 | 1221.56 | 3.202360 | 2.87839 | 1345.87 |
| 0.08 | 2.990789 | 2.79167 | 1215.11 | 3.216545 | 2.88932 | 1357.73 |
| 0.09 | 2.977018 | 2.78971 | 1208.73 | 3.230817 | 2.90046 | 1369.81 |
| 0.1 | 2.963370 | 2.78774 | 1202.41 | 3.245176 | 2.91179 | 1382.12 |
| 0.2 | 2.833462 | 2.76767 | 1142.76 | 3.392920 | 3.03827 | 1520.14 |
| 0.3 | 2.714550 | 2.74721 | 1088.85 | 3.544203 | 3.19707 | 1693.42 |

## IV. APPLICATION TO THE LENSED QUASAR SDSS J1004 + 4112

The system SDSS J1004 + 4112 discovered in the Sloan Digital Sky Survey has a background source at a redshift $z_S = 1.734$, identified as a quasar lensed into five images, $A$, $B$, $C$, $D$, and $E$, by an intervening galaxy cluster at a redshift $z_L = 0.68$ [33–37]. The galaxy cluster is assumed to have a spherical inner structure, despite the fact that the presence of five images obviously violates this assumption. We find it convenient to express the observed position angles of images with respect to the galaxy cluster as

$$\alpha_A = 8''.37 \pm 0''.04, \quad \alpha_B = 8''.48 \pm 0''.04,$$
$$\alpha_C = 9''.73 \pm 0''.04, \quad \alpha_D = 5''.45 \pm 0''.04,$$
$$\alpha_E = 0''.21 \pm 0''.04, \tag{65}$$

where we have taken the same position angle error used in Refs. [38,39].

Throughout the application, we adopt, for the Hubble constant and the present dark energy density, a $\Lambda$CDM model's best-fit values provided by Planck collaboration [6]—$H_0 = (67.36 \pm 0.54) \text{ km s}^{-1} \text{ Mpc}^{-1}$ and $\Omega_{\Lambda 0} = 0.6847 \pm 0.0073$—from which we deduce the cosmological constant value, $\Lambda = (1.08914 \pm 0.02907) \times 10^{-52} \text{m}^{-2}$. Accordingly, the present matter and the present curvature densities are related by $\Omega_{\rho 0} = 0.3153 - \Omega_{k0}$.

### A. Flat universe ($k = 0$)

We start by addressing the case without spatial curvature in order, inter alia, to appreciate discrepancies when the nonflat spaces will be considered afterward. The calculation of the source position angle $-\varphi_S$ using (55) or (56), and the time delay $\Delta t$ using (63), for the image pairs $(A, B)$, $(A, C)$, $(D, A)$, $(B, C)$, and $(D, C)$, are carried out by fitting the galaxy cluster mass in function of maximum $+$, central $\pm 0$, and minimum $-$ values of $\Omega_{\Lambda 0}$, $\alpha$, and $\alpha'$. For convenience, the present scale factor $a_0$ is normalized to unity.[2] The results are displayed in Table I, where only some specific combinations of $(\Omega_{\Lambda 0}, \alpha, \alpha')$ within their error bars have been shown, corresponding to the upper, central, and lower limit values of $M$, $-\varphi_S$, and $\Delta t$, which are sufficient for estimating their uncertainties summarized in Table II.

TABLE VII. Variation of the galaxy cluster mass $M$, the position angle $-\varphi_S$, and the time delay $\Delta t$ versus the present curvature density $\Omega_{k0}$ within the range $[-0.3, 0.3]$ for the observed image pair $(D, A)$ of the lensed quasar SDSS J1004 + 4112 in curved Einstein-Straus–de Sitter space-time ($k = \pm 1$). The angles $\alpha_D$ and $\alpha_A$ as well as the present dark energy density $\Omega_{\Lambda 0}$ are fixed in their central values.

| $(D, A)$ | $k = +1$ | | | $k = -1$ | | |
|---|---|---|---|---|---|---|
| $\|\Omega_{k0}\|$ | $M[10^{13} M_\odot]$ | $-\varphi_S['']$ | $\Delta t$[days] | $M[10^{13} M_\odot]$ | $-\varphi_S['']$ | $\Delta t$[days] |
| 0.0001 | 1.740161 | 6.02817 | 2049.34 | 1.740321 | 6.02829 | 2049.58 |
| 0.0002 | 1.740078 | 6.02813 | 2049.22 | 1.740398 | 6.02837 | 2049.70 |
| 0.0003 | 1.739994 | 6.02809 | 2049.11 | 1.740475 | 6.02846 | 2049.83 |
| 0.0004 | 1.739911 | 6.02805 | 2048.99 | 1.740552 | 6.02854 | 2049.95 |
| 0.0005 | 1.739827 | 6.02801 | 2048.88 | 1.740628 | 6.02862 | 2050.08 |
| 0.0006 | 1.739744 | 6.02797 | 2048.76 | 1.740705 | 6.02870 | 2050.21 |
| 0.0007 | 1.739660 | 6.02793 | 2048.65 | 1.740782 | 6.02879 | 2050.33 |
| 0.0008 | 1.739577 | 6.02789 | 2048.54 | 1.740859 | 6.02887 | 2050.46 |
| 0.0009 | 1.739493 | 6.02785 | 2048.42 | 1.740936 | 6.02895 | 2050.58 |
| 0.001 | 1.739410 | 6.02781 | 2048.31 | 1.741012 | 6.02904 | 2050.71 |
| 0.002 | 1.738577 | 6.02741 | 2047.16 | 1.741781 | 6.02987 | 2051.97 |
| 0.003 | 1.737744 | 6.02702 | 2046.02 | 1.742550 | 6.03070 | 2053.23 |
| 0.004 | 1.736912 | 6.02662 | 2044.88 | 1.743320 | 6.03153 | 2054.49 |
| 0.005 | 1.736081 | 6.02622 | 2043.74 | 1.744090 | 6.03236 | 2055.76 |
| 0.006 | 1.735250 | 6.02583 | 2042.61 | 1.744861 | 6.03319 | 2057.02 |
| 0.007 | 1.734421 | 6.02543 | 2041.47 | 1.745632 | 6.03402 | 2058.29 |
| 0.008 | 1.733591 | 6.02503 | 2040.33 | 1.746404 | 6.03486 | 2059.56 |
| 0.009 | 1.732763 | 6.02463 | 2039.20 | 1.747177 | 6.03569 | 2060.83 |
| 0.01 | 1.731936 | 6.02423 | 2038.07 | 1.747950 | 6.03653 | 2062.11 |
| 0.02 | 1.723704 | 6.02022 | 2026.81 | 1.755712 | 6.04490 | 2074.93 |
| 0.03 | 1.715547 | 6.01617 | 2015.68 | 1.763531 | 6.05334 | 2087.92 |
| 0.04 | 1.707465 | 6.01209 | 2004.67 | 1.771406 | 6.06183 | 2101.09 |
| 0.05 | 1.699457 | 6.00799 | 1993.78 | 1.779338 | 6.07038 | 2114.43 |
| 0.06 | 1.691523 | 6.00385 | 1983.01 | 1.787325 | 6.07899 | 2127.96 |
| 0.07 | 1.683661 | 5.99969 | 1972.36 | 1.795368 | 6.08765 | 2141.67 |
| 0.08 | 1.675870 | 5.99550 | 1961.82 | 1.803466 | 6.09638 | 2155.57 |
| 0.09 | 1.668151 | 5.99129 | 1951.40 | 1.811618 | 6.10517 | 2169.67 |
| 0.1 | 1.660502 | 5.98706 | 1941.09 | 1.819824 | 6.11403 | 2183.96 |
| 0.2 | 1.587690 | 5.94388 | 1843.74 | 1.904550 | 6.20611 | 2338.84 |
| 0.3 | 1.521045 | 5.89987 | 1755.85 | 1.992164 | 6.30499 | 2519.23 |

At first sight, the Table II indicates that a small image angular difference $\Delta \alpha$ leads to a small polar angle $-\varphi_S$ as well as a small time delay and vice versa. This offers remarkably two important consequences. First, it concords with our theoretical model which states that $-\varphi_S$ and $\Delta t$ vanish for $\alpha = \alpha'$. Second, the reported time delays obey the temporal ordering sequence C-B-A-D, where the longest and shortest time delays correspond to the longest and shortest angular differences $\Delta \alpha$—the more widely separated $C$ and $D$ images and the more closely separated $A$ and $B$ images—respectively. This is all to say that a larger $\Delta \alpha$ leads to a larger $-\varphi_S$ as well as a longer $\Delta t$. The angular difference has then the same implications on the time delay hierarchy as it would be the case for the commonly used image angular separation. In particular, the broad range of time delays provides lower and upper bound estimates on the galaxy cluster mass, $1.717 \times 10^{13} M_\odot \leq M \leq 3.177 \times 10^{13} M_\odot$, or $M = (2.447 \pm 0.73) \times 10^{13} M_\odot$, which matches perfectly with the enclosed mass within a radius of 60 kpc, measured by Williams et al. ($\sim 2.5 \times 10^{13} M_\odot$) with Chandra x-ray observations, but roughly twice as small as their measurements within 100 kpc ($\sim 5 \times 10^{13} M_\odot$) [35,39–41]. Additionally, the time delay between $A$ and $B$ images is affected by a margin of uncertainty that is large enough to allow a plausible comparison with the results of Kawano and Oguri [37], $\Delta t_{AB} \lesssim 28$ days, and Fohlmeister et al. [36] $\Delta t_{AB} = (40.6 \pm 1.8)$ days. Our predicted time delays between $B$ and $C$, and between $D$ and $C$ are further consistent with what Kawano and Oguri have obtained, $\Delta t_{BC} \lesssim 1400$ days, and $\Delta t_{DC} \lesssim 3700$ days. Fohlmeister et al. have used a mass model to predict a time delay of approximately 2000 days between $D$ and $A$, in agreement with our result. The same authors have measured relative to the image pairs $(B, C)$ and $(A, C)$, the following time delays: $\Delta t_{BC} = (782 \pm 7)$ days, $\Delta t_{AC} = (821.6 \pm 2.1)$ days, which are nonetheless quite shorter than our predictions. Likewise for Liu et al. who have lately reported in Ref. [42] time delays that differ from ours, but coincide with those of Muñoz et al. [43] and recent measurements by Perera et al. [41]: $\Delta t_{BC} = (781.92 \pm 2.20)$ days, $\Delta t_{AC} = (825.99 \pm 2.10)$ days, and $\Delta t_{DC} = (2456.99 \pm 5.55)$ days.

---

[2]In the flat Universe, the present scale factor $a_0$ can be set arbitrarily without loss of generality.

TABLE VIII. Variation of the galaxy cluster mass $M$, the position angle $-\varphi_S$, and the time delay $\Delta t$ versus the present curvature density $\Omega_{k0}$ within the range $[-0.3, 0.3]$ for the observed image pair $(D, C)$ of the lensed quasar SDSS J1004 + 4112 in curved Einstein-Straus–de Sitter space-time ($k = \pm 1$). The angles $\alpha_D$ and $\alpha_C$ as well as the present dark energy density $\Omega_{\Lambda 0}$ are fixed in their central values.

| $(D, C)$ | $k = +1$ | | | $k = -1$ | | |
|---|---|---|---|---|---|---|
| $|\Omega_{k0}|$ | $M[10^{13} M_\odot]$ | $-\varphi_S['']$ | $\Delta t$[days] | $M[10^{13} M_\odot]$ | $-\varphi_S['']$ | $\Delta t$[days] |
| 0.0001 | 2.022659 | 8.83494 | 3249.92 | 2.022846 | 8.8351 | 3250.3 |
| 0.0002 | 2.022562 | 8.83488 | 3249.74 | 2.022936 | 8.83519 | 3250.49 |
| 0.0003 | 2.022465 | 8.83483 | 3249.56 | 2.023026 | 8.83529 | 3250.69 |
| 0.0004 | 2.022368 | 8.83477 | 3249.37 | 2.023116 | 8.83539 | 3250.88 |
| 0.0005 | 2.022271 | 8.83471 | 3249.19 | 2.023206 | 8.83549 | 3251.07 |
| 0.0006 | 2.022174 | 8.83465 | 3249.00 | 2.023296 | 8.83558 | 3251.27 |
| 0.0007 | 2.022077 | 8.83459 | 3248.82 | 2.023386 | 8.83568 | 3251.46 |
| 0.0008 | 2.021981 | 8.83454 | 3248.64 | 2.023476 | 8.83578 | 3251.65 |
| 0.0009 | 2.021884 | 8.83448 | 3248.45 | 2.023566 | 8.83588 | 3251.85 |
| 0.001 | 2.021787 | 8.83442 | 3248.27 | 2.023656 | 8.83597 | 3252.04 |
| 0.002 | 2.020818 | 8.83384 | 3246.44 | 2.024556 | 8.83695 | 3253.97 |
| 0.003 | 2.019850 | 8.83326 | 3244.60 | 2.025457 | 8.83792 | 3255.91 |
| 0.004 | 2.018883 | 8.83268 | 3242.78 | 2.026359 | 8.83890 | 3257.85 |
| 0.005 | 2.017917 | 8.8321 | 3240.95 | 2.027261 | 8.83987 | 3259.79 |
| 0.006 | 2.016952 | 8.83151 | 3239.12 | 2.028164 | 8.84085 | 3261.74 |
| 0.007 | 2.015987 | 8.83093 | 3237.30 | 2.029068 | 8.84183 | 3263.68 |
| 0.008 | 2.015024 | 8.83035 | 3235.48 | 2.029972 | 8.8428 | 3265.63 |
| 0.009 | 2.014061 | 8.82976 | 3233.66 | 2.030877 | 8.84378 | 3267.58 |
| 0.01 | 2.013100 | 8.82918 | 3231.84 | 2.031783 | 8.84476 | 3269.54 |
| 0.02 | 2.003532 | 8.82330 | 3213.78 | 2.040877 | 8.85454 | 3289.21 |
| 0.03 | 1.994052 | 8.81737 | 3195.92 | 2.050039 | 8.86435 | 3309.14 |
| 0.04 | 1.984659 | 8.81140 | 3178.26 | 2.059270 | 8.87419 | 3329.31 |
| 0.05 | 1.975351 | 8.80538 | 3160.8 | 2.068568 | 8.88405 | 3349.75 |
| 0.06 | 1.966129 | 8.79932 | 3143.52 | 2.077934 | 8.89394 | 3370.44 |
| 0.07 | 1.956992 | 8.79322 | 3126.44 | 2.087367 | 8.90385 | 3391.41 |
| 0.08 | 1.947937 | 8.78709 | 3109.55 | 2.096867 | 8.91378 | 3412.64 |
| 0.09 | 1.938965 | 8.78092 | 3092.84 | 2.106432 | 8.92373 | 3434.16 |
| 0.1 | 1.930074 | 8.77472 | 3076.30 | 2.116063 | 8.93369 | 3455.96 |
| 0.2 | 1.845448 | 8.71147 | 2920.35 | 2.215671 | 9.03380 | 3690.85 |
| 0.3 | 1.767987 | 8.64699 | 2779.69 | 2.319147 | 9.13166 | 3960.63 |

Failing to answer the question of what and how much is the effect of a positive $\Lambda$ on the bending of light, we have made Table III, based upon the variation of the cosmological constant $\Lambda$ within its error bar, from the lower value to the upper one, with keeping $\alpha$ and $\alpha'$ fixed in their central values. Quantitatively, all parameters are related monotonously to each other, but an increase of $\Lambda$ by 5.34% leads to a relatively slight decrease of $-\varphi_S$ by 1.32% associated though with a very slightly increasing mass by 0.27%. The same values hold for all the image pairs. Even though the method is somewhat less robust than desired, the effect appears to be very small on cosmological scales. Note also that the time delay gets a relatively very slight increase as well: 0.44% for the image pairs $(A, B)$, $(BC)$, and $(A, C)$ and 0.46% for $(D, A)$, and 0.47% for $(D, C)$, which are almost the same. While this statement has only been made in the framework of flat ESdS model, Hu et al. [29] have

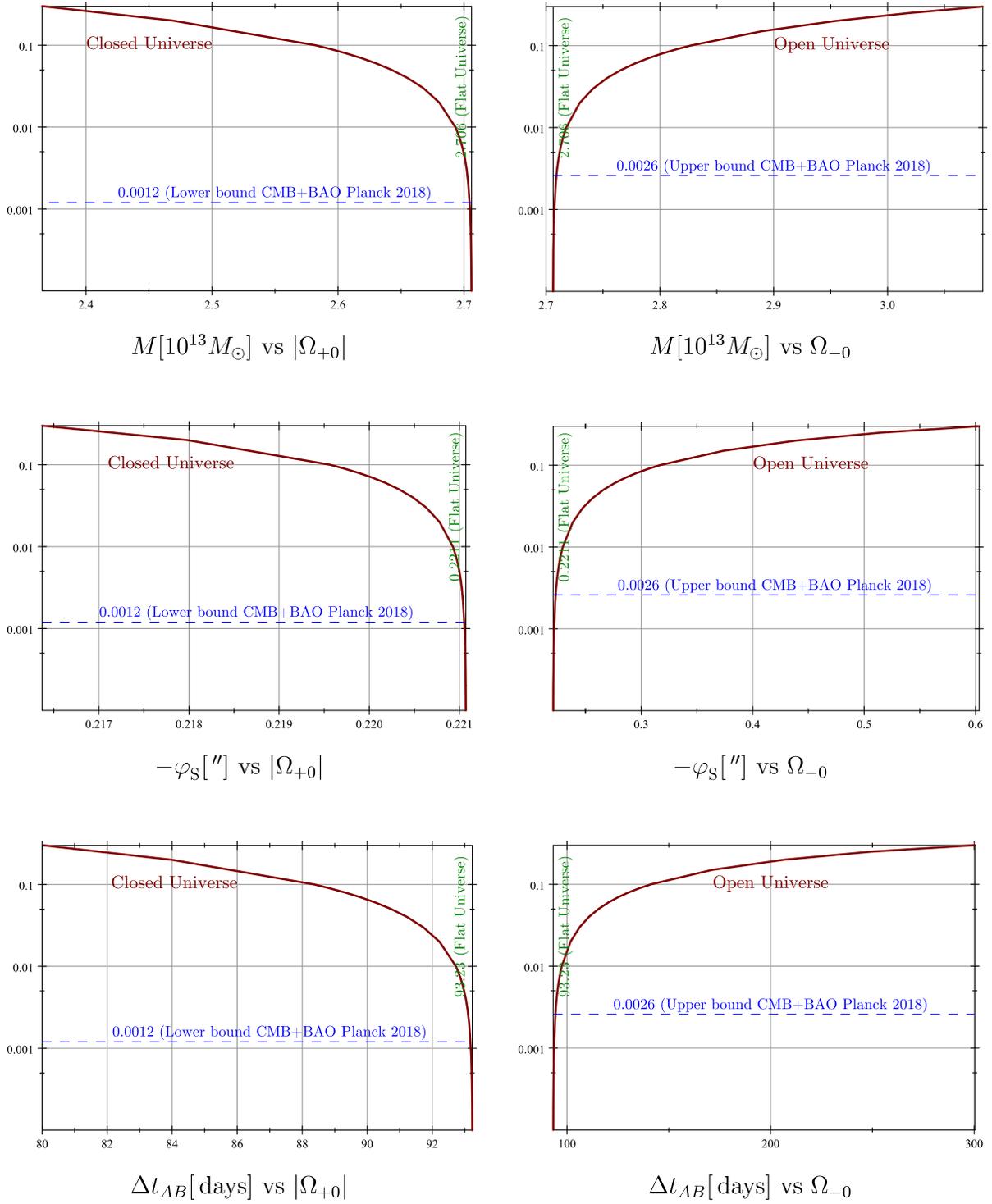

FIG. 2. Evolution of the galaxy cluster mass $M$, the position angle $-\varphi_S$, and the time delay $\Delta t$ versus the present curvature density $\Omega_{k0}$ within the range $[-0.3, 0.3]$ for the observed image pair $(A, B)$ of the lensed quasar SDSS J1004 + 4112 in curved Einstein-Straus–de Sitter space-time ($k = \pm 1$). The angles $\alpha_A$ and $\alpha_B$ as well as the present dark energy density $\Omega_{\Lambda 0}$ are fixed in their central values.

proceeded otherwise and more rigorously using a curved ESdS model in a simple case where the source is not inclined but aligned with both the lens and the observer. They isolate all affects that can be produced by other parameters on the bending of light when varying $\Lambda$. Especially, they fix the angular diameter distance as well as the mass $M$ and the radius of the vacuole by compensating the changes in $\Omega_{\Lambda 0}$ by $\Omega_{k0}$. The calculations were

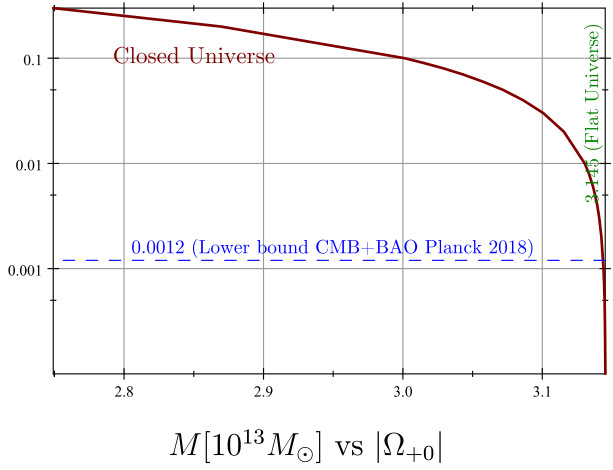
$M[10^{13}M_\odot]$ vs $|\Omega_{+0}|$

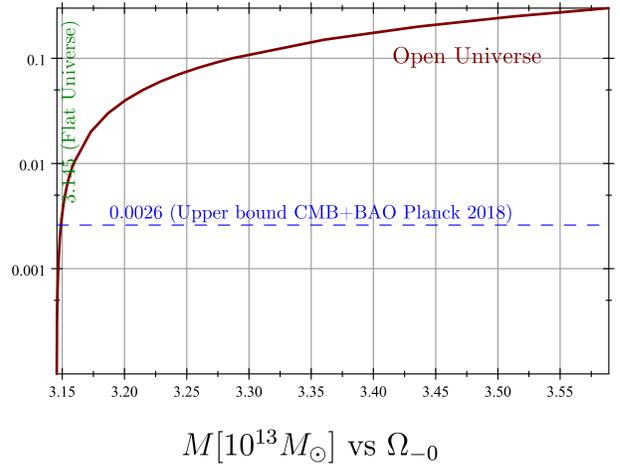
$M[10^{13}M_\odot]$ vs $\Omega_{-0}$

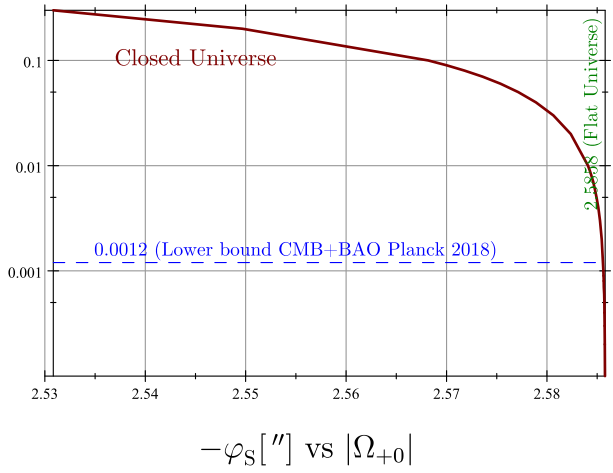
$-\varphi_S['']$ vs $|\Omega_{+0}|$

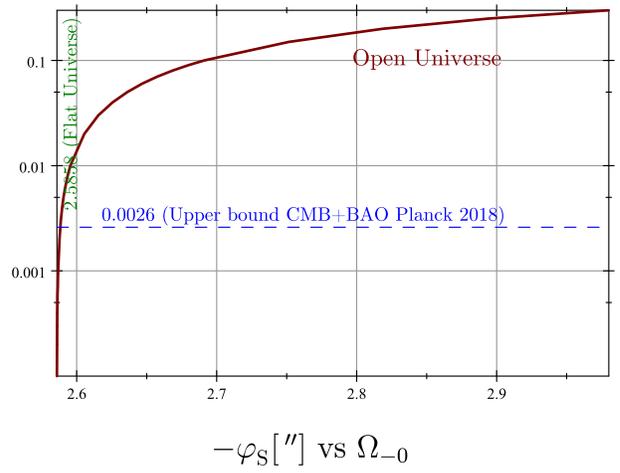
$-\varphi_S['']$ vs $\Omega_{-0}$

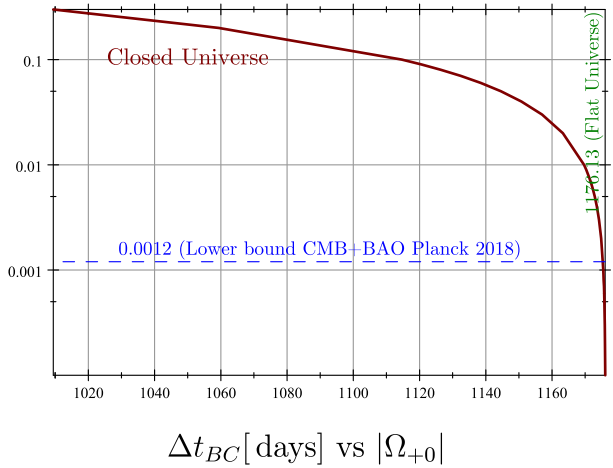
$\Delta t_{BC}[\,\text{days}\,]$ vs $|\Omega_{+0}|$

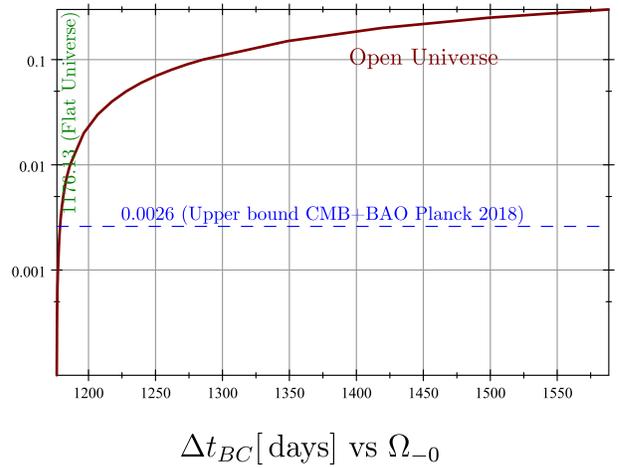
$\Delta t_{BC}[\,\text{days}\,]$ vs $\Omega_{-0}$

FIG. 3. Evolution of the galaxy cluster mass $M$, the position angle $-\varphi_S$, and the time delay $\Delta t$ versus the present curvature density $\Omega_{k0}$ within the range $[-0.3, 0.3]$ for the observed image pair $(B, C)$ of the lensed quasar SDSS J1004 + 4112 in curved Einstein-Straus–de Sitter space-time ($k = \pm 1$). The angles $\alpha_B$ and $\alpha_C$ as well as the present dark energy density $\Omega_{\Lambda 0}$ are fixed in their central values.

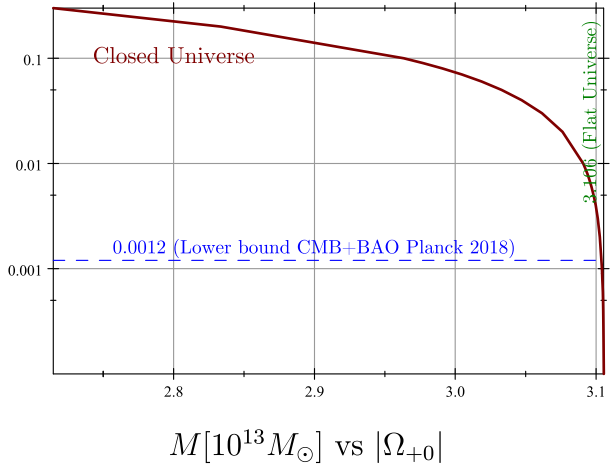
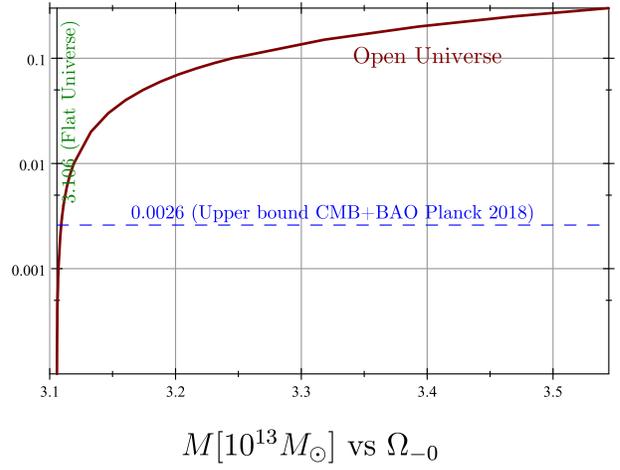

$M[10^{13}M_\odot]$ vs $|\Omega_{+0}|$      $M[10^{13}M_\odot]$ vs $\Omega_{-0}$

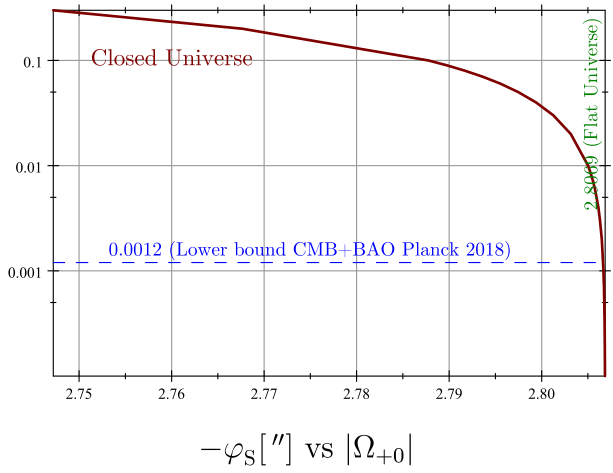
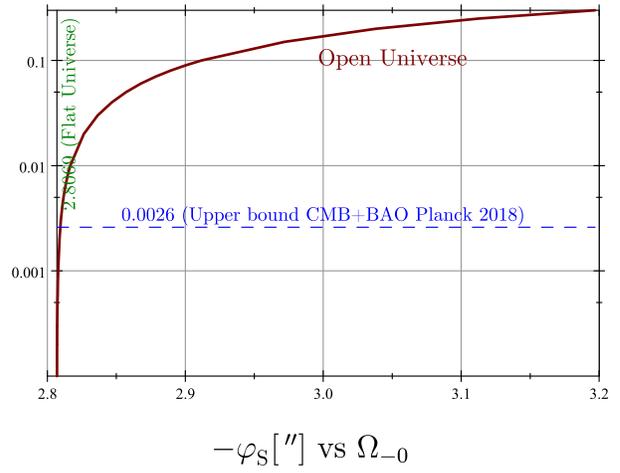

$-\varphi_\mathrm{S}['']$ vs $|\Omega_{+0}|$      $-\varphi_\mathrm{S}['']$ vs $\Omega_{-0}$

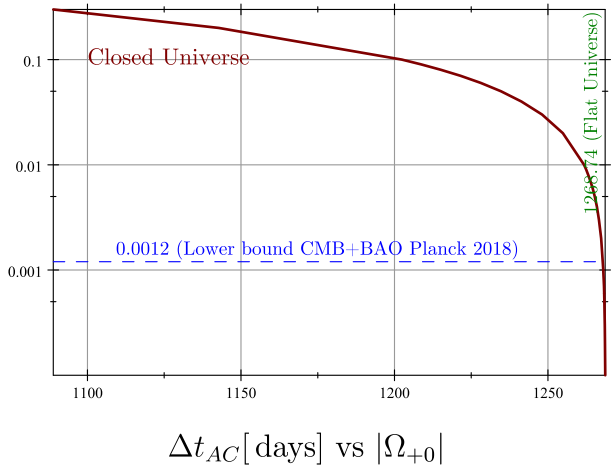
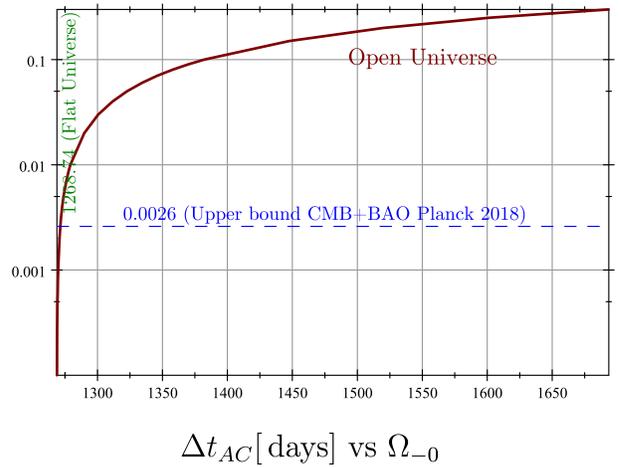

$\Delta t_{AC}[\,\mathrm{days}\,]$ vs $|\Omega_{+0}|$      $\Delta t_{AC}[\,\mathrm{days}\,]$ vs $\Omega_{-0}$

FIG. 4. Evolution of the galaxy cluster mass $M$, the position angle $-\varphi_\mathrm{S}$, and the time delay $\Delta t$ versus the present curvature density $\Omega_{k0}$ within the range $[-0.3, 0.3]$ for the observed image pair $(A, C)$ of the lensed quasar SDSS J1004 + 4112 in curved Einstein-Straus–de Sitter space-time ($k = \pm 1$). The angles $\alpha_A$ and $\alpha_C$ as well as the present dark energy density $\Omega_{\Lambda 0}$ are fixed in their central values.

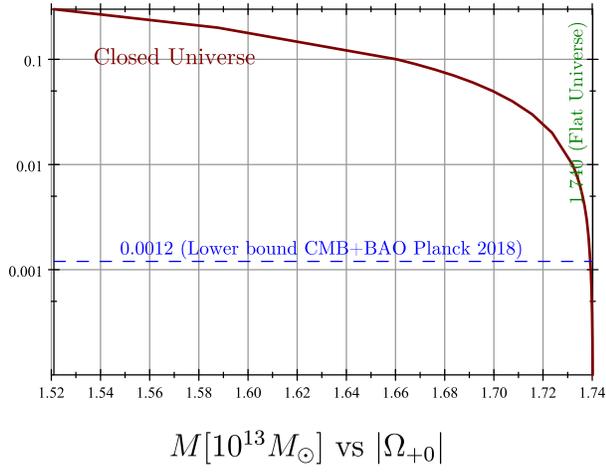
$M[10^{13} M_\odot]$ vs $|\Omega_{+0}|$

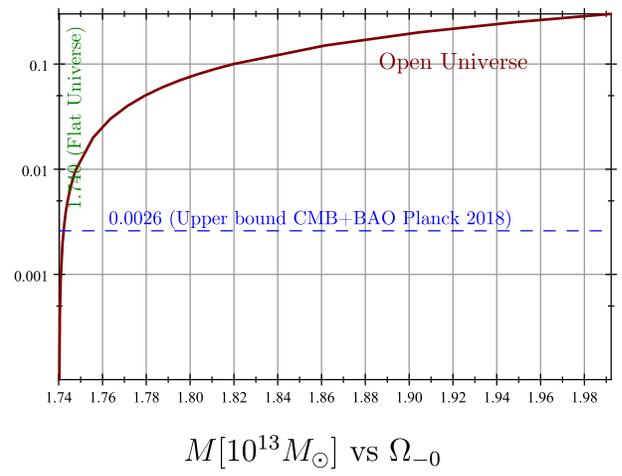
$M[10^{13} M_\odot]$ vs $\Omega_{-0}$

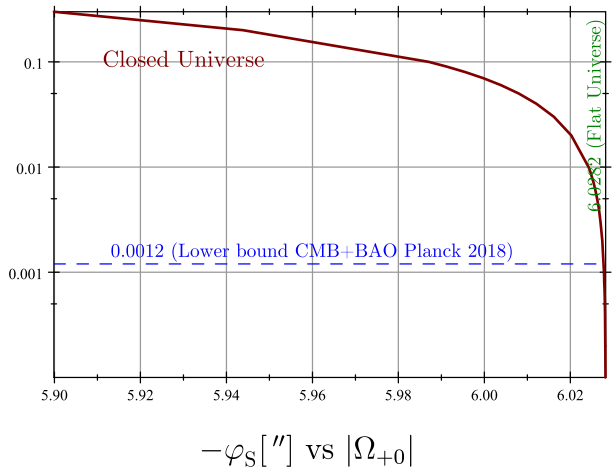
$-\varphi_S['']$ vs $|\Omega_{+0}|$

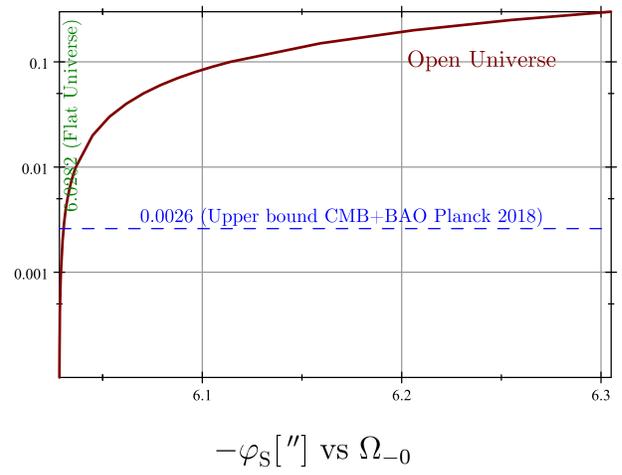
$-\varphi_S['']$ vs $\Omega_{-0}$

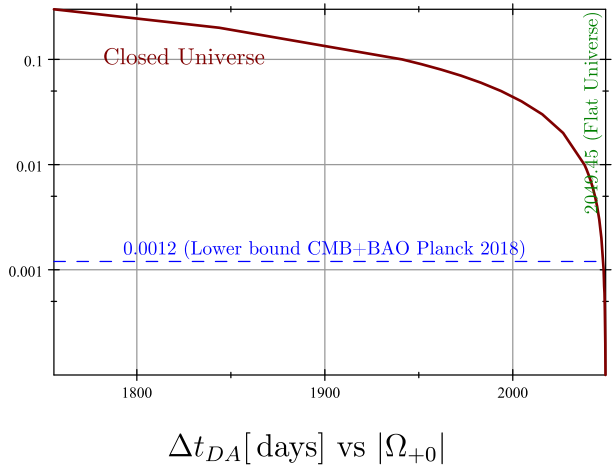
$\Delta t_{DA}[\text{days}]$ vs $|\Omega_{+0}|$

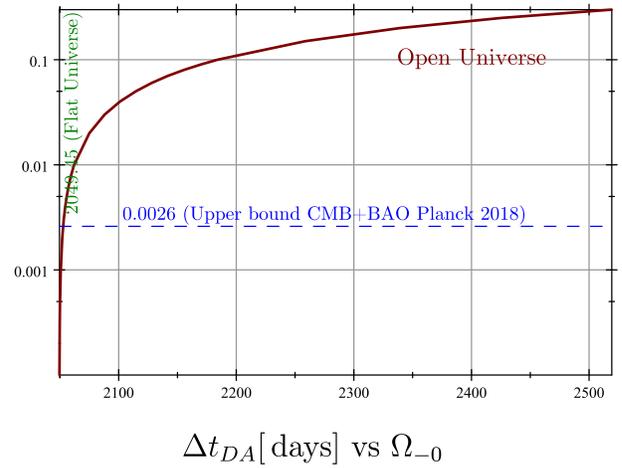
$\Delta t_{DA}[\text{days}]$ vs $\Omega_{-0}$

FIG. 5. Evolution of the galaxy cluster mass $M$, the position angle $-\varphi_S$, and the time delay $\Delta t$ versus the present curvature density $\Omega_{k0}$ within the range $[-0.3, 0.3]$ for the observed image pair $(D, A)$ of the lensed quasar SDSS J1004 + 4112 in curved Einstein-Straus–de Sitter space-time ($k = \pm 1$). The angles $\alpha_D$ and $\alpha_A$ as well as the present dark energy density $\Omega_{\Lambda 0}$ are fixed in their central values.

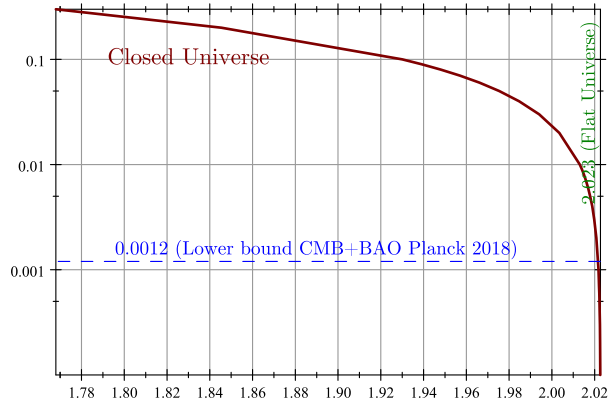
$M[10^{13} M_\odot]$ vs $|\Omega_{+0}|$

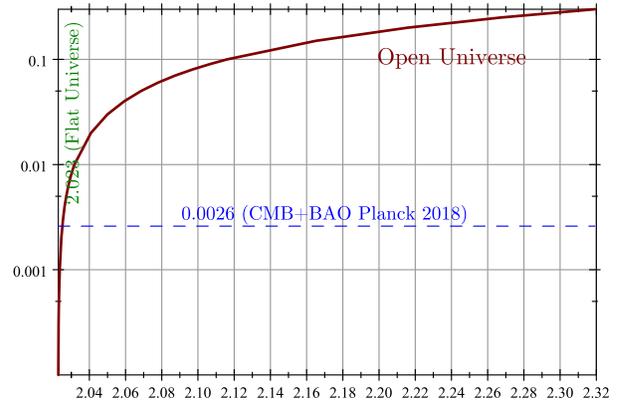
$M[10^{13} M_\odot]$ vs $\Omega_{-0}$

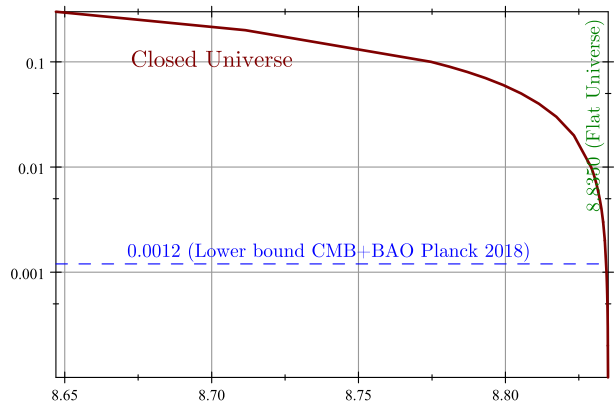
$-\varphi_S['']$ vs $|\Omega_{+0}|$

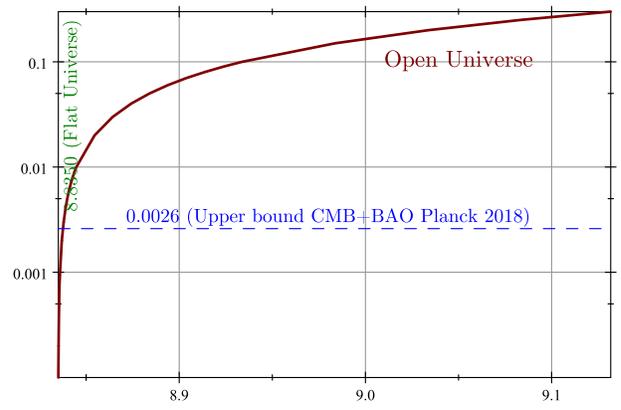
$-\varphi_S['']$ vs $\Omega_{-0}$

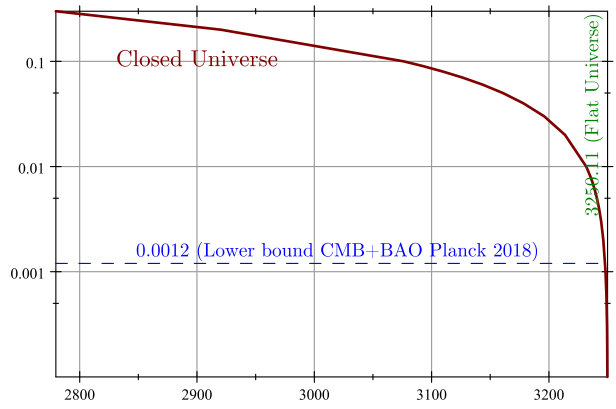
$\Delta t_{DC}[\text{days}]$ vs $|\Omega_{+0}|$

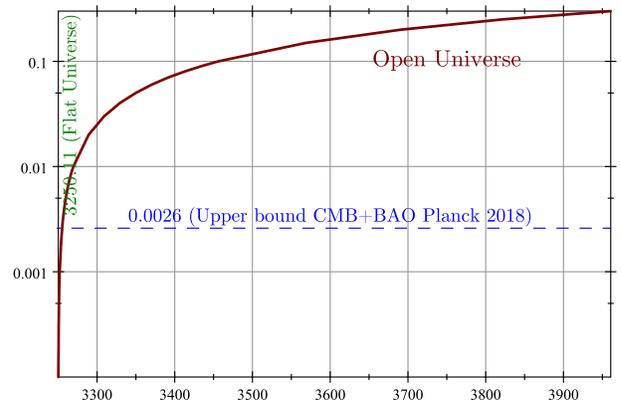
$\Delta t_{DC}[\text{days}]$ vs $\Omega_{-0}$

FIG. 6. Evolution of the galaxy cluster mass $M$, the position angle $-\varphi_S$, and the time delay $\Delta t$ versus the present curvature density $\Omega_{k0}$ within the range $[-0.3, 0.3]$ for the observed image pair $(D, C)$ of the lensed quasar SDSS J1004 + 4112 in curved Einstein-Straus–de Sitter space-time ($k = \pm 1$). The angles $\alpha_D$ and $\alpha_C$ as well as the present dark energy density $\Omega_{\Lambda 0}$ are fixed in their central values.

performed for different masses and showed that the Λ effect on light bending is very tiny and can be neglected.

### B. Positively curved universe ($k = +1$)

Now, we are going to study the effect of a positive spatial curvature on light deflection and time delay in the framework of a closed ESdS model. A better way to do this is to let the present curvature density $\Omega_{+0}$ free to vary discretely upon a set of $(\Omega_{\Lambda 0}, \alpha, \alpha')$ fixed in their central values. Thereupon, we fit the galaxy cluster mass for various values of $\Omega_{+0}$, within the range $[-0.3, -0.0001]$,[3] to calculate the polar angle $-\varphi_S$ and the time delay $\Delta t$ of the four image pairs. The results are quoted in the Tables IV, V, VI, VII, and VIII. It would also be best to make use of plots, by interpolating the data points $(M, |\Omega_{+0}|)$, $(-\varphi_S, |\Omega_{+0}|)$, and $(\Delta t, |\Omega_{+0}|)$ separately. We note that inserting the logarithm function onto the y axis allows better visualization of the real effect in the vicinity of smaller $|\Omega_{+0}|$.

According to the Tables IV, V, VI, VII, and VIII, as well as the Figs. 2, 3, 4, 5, and 6, the galaxy cluster mass $M$, the polar angle $-\varphi_S$ as well as the time delay $\Delta t$ follow the same behavior against the present curvature density $|\Omega_{+0}|$. Specifically, they decrease significantly upon increasing $|\Omega_{+0}|$ within its larger values. This is what the downward sloping part of graphs demonstrates. However, for smaller $|\Omega_{+0}|$, the parameters increase very slightly until they become steady. These limiting values through smaller $|\Omega_{+0}|$ are none other than those of the flat ESdS model in the Table I. These features remain the same regardless of which values are chosen for $\Omega_{\Lambda 0}, \alpha$, and $\alpha'$ within their error bars. An added feature that is easy to check would be that the parameters $M, -\varphi_S$, and $\Delta t$ are correlated almost linearly to each other through all the range of $|\Omega_{+0}|$. In view of this finding, we conclude that a small curvature density does not significantly impact the light deflection and the time delay.

### C. Negatively curved universe ($k = -1$)

Analogously, we allow several values of $\Omega_{-0}$ within the range $[0.0001, 0.3]$ to show its impact on $M, -\varphi_S$ and $\Delta t$ in the hyperbolic ESdS space-time. The results are reported in Tables IV, V, VI, VII, and VIII, and graphically presented in Figs. 2, 3, 4, 5, and 6. Conversely to the closed ESdS model, it is easy to see that the three parameters grow significantly upon increasing $\Omega_{-0}$ beyond its smaller values (the upward sloping part of graphs), whereas they decrease very slightly within smaller $\Omega_{-0}$, until they become steady once reaching the parameter values of the flat ESdS model in the Table I. In this context one should note that this reverse effect is quite expected since the spatial curvature has changed its sign. Again, one can easily check that the relationship between the parameters are approximately described by a linear function through all the range of $\Omega_{-0}$. Finally, we have come to the same conclusion as for the closed Universe: including a small spatial curvature in the ESdS metric does not appreciably affect the light deflection and time delay.

## V. CONCLUSION

In this work, we have generalized the computation of light deflection and time delay in the Einstein-Straus–de Sitter space-time, by modeling the gravitational lens as a static Schwarzschild–de Sitter vacuole embedded in an external FLRW Universe, with the purpose of covering together all three types of spatial curvature, $k = 0$ (flat Universe), $k = +1$ (closed Universe), and $k = -1$ (open Universe). This study results in generalized analytical expressions for the light deflection and the time delay.

After that, numerical applications to the lensed quasar SDSS J1004 + 4112 have been thoroughly performed for each case separately. Assuming first the flat Einstein-Straus–de Sitter background, predictions of five time delays between the four bright images of the aforementioned lensed quasar have been obtained and compared to some measurements in other research, $(3250 \pm 64)$ days for the well-known longest time delay between the images $D$ and $C$, and $(93 \pm 70)$ days for the shortest one between the images $A$ and $B$. For the other three time delays, we have gotten $2049^{+59}_{-58}$ days between the images $D$ and $A$, $(1269 \pm 77)$ days between the images $A$ and $C$, and $1176^{+78}_{-77}$ days between the images $B$ and $C$. These time delays follow from the fit of the galaxy cluster mass for each image pair independently. The average mass that we have found is estimated to be $2.447 \times 10^{13} M_\odot$, with an accuracy of about 30%.

Additionally, we have shortly discussed the relationship between the light deflection and the cosmological constant and reached the following conclusion. Although the light deflection is attenuated in the presence of a positive cosmological constant, the results apparently seem to tell us that the effect is numerically less significant on cosmological scales, as has been recently asserted by Hu *et al.* [29]. The same conclusion holds for the time delay [11]. As for whether the instruments could observe this tiny effect, that is, in our view, another mission much more difficult.

After that, we have successively considered the Einstein-Straus–de Sitter model with positive and negative spatial curvature. A discrete variation of the present curvature density in the range $[-0.3, 0.3]$ is permitted to investigate its contribution to three parameters, the galaxy cluster mass, the light deflection, as well as the time delay. The results have been reported in tables, and graphically shown in figures, from which a number of conclusions have been drawn. Although the three parameters are linearly influenced by each other, assigning larger values to the present curvature density will impact them significantly. Clearly, the parameters are not subject to the same dependence on the modulus of the present curvature density $|\Omega_{k0}|$ as it increases; they are decreasing in the closed Universe whereas they are increasing in the open Universe. This reversing effect is expected for cosmological models with different signs of curvature. However, the effect thereof is increasingly disappearing when the curvature density gets smaller and smaller $|\Omega_{k0}| \lessapprox 0.001$, covering the parameters of the flat Universe. This characteristic holds true for both closed and open Einstein-Straus–de Sitter models. The precise value of the curvature density is currently still under active investigation, but if the observations truly trend in favor of small values thereof, one may feel confident saying that the expected small curvature of the current Universe is, as it were, not needed for the computation of light deflection and time delay in the Einstein-Straus–de Sitter model. On the other hand, in the cases where the present curvature density becomes important, this will profoundly influence the strong lensing phenomenon. These results indicate the robustness of our generalized approach, and, more importantly, they allow us to accurately constrain the spatial curvature on the basis of observational light deflection and time delay data.

---

[3]According to (5), the present curvature density is constrained by $\Omega_{k0} < 0.3153$, since the present matter density $\Omega_{\rho 0}$ must be positive definite.

To our knowledge, very little research tackles the question of $\Omega_{k0}$ effect on gravitational lensing. The author of Ref. [44] exploits weak-lensing data to constrain the curvature density on which the angular diameter distance between the lens and the source depends. A similar study is carried out in Ref. [45]. In Ref. [46], it was shown that the supernova light-curve parameters are unaffected by the curvature density. In Ref. [47], a closed Universe has been considered to analyze the evolution of the dark energy and matter densities against $\Omega_{k0}$, demonstrating that the only conceivable values correspond to small curvature densities. The dependence of the gravitational potential on the spatial curvature in the weak field limit has been proved in Ref. [48], using the same cosmological parameters as ours with realistic small curvature densities [6]. Finally, in Ref. [49], the authors have derived a particular local static FLRW metric from the osculating de Sitter metric, concluding that the spatial curvature effects on the local dynamics are minor but may not be negligible.

Recently in Ref. [41], Perera *et al.* have estimated a shorter time delay of roughly 8 yr between the image $C$ and the faint fifth image $E$ [50], very close to what Forés-Toribio *et al.* have obtained before [39], and a longer one of roughly 9 yr. We have overlooked the image $E$ in this work owing to the fact that fitting the cluster mass is not possible or yields values that are far from acceptable. The image in question is located very close to the center of the galaxy cluster (∼0.2 arcsec) (65), meaning that the associated photon inevitably encounters the mass distribution along the path. While it may depend on the nonspherical mass distribution, to which the photon is more sensitive, we think including an interior Schwarzschid–de Sitter solution [51] to the Einstein-Straus–de Sitter model would likely be a possible route to circumvent this problem. We are currently challenging ourselves to respond to this question in the hope it will be issued in the near future. Another avenue to test whether our findings are reliable would be to expand our analysis to cover other lensed quasar systems.